\documentclass[referee,a4paper,12pt,traditabstract]{swsc} 

\usepackage{graphicx}
\usepackage{amsmath}
\usepackage{textcomp}
\usepackage{txfonts}
\usepackage{subfigure}
\usepackage{lineno}
\usepackage[authoryear,round]{natbib}
\usepackage[backref]{hyperref}
\usepackage{url}
\usepackage{array}
\usepackage{enumerate}
\usepackage{longtable}
\usepackage{multirow}
\usepackage{float}

\hypersetup{colorlinks=true,citecolor=cyan,urlcolor=cyan,linkcolor=blue}

\begin{document}

%\begin{linenumbers}

   \title{Catalogue of 55\textendash 80 MeV Solar Proton Events Extending Through Solar Cycles 23 and 24}
   
   \titlerunning{Catalogue of 55\textendash 80 MeV Solar Proton Events}

   \authorrunning{Paassilta et al.}

   \author{Miikka Paassilta
          \inst{1}\fnmsep\thanks{Corresponding author, \email{\href{mailto:mimapa@utu.fi} mimapa@utu.fi}}
\and
Osku Raukunen\inst{1}
\and
Rami Vainio\inst{1}
\and
Eino Valtonen\inst{1}
\and
Athanasios Papaioannou\inst{2}
\and
Robert Siipola\inst{1}
\and
Esa Riihonen\inst{1}        
\and
Mark Dierckxsens\inst{3}
\and
Norma Crosby\inst{3}
\and
Olga Malandraki\inst{2}
\and
Bernd Heber\inst{4}
\and
Karl-Ludwig Klein\inst{5}}

\institute{Department of Physics and Astronomy, University of Turku, 20014 Finland
\and
Institute for Astronomy, Astrophysics, Space Applications and Remote Sensing (IAASARS), National Observatory of Athens, I. Metaxa \& Vas. Pavlou St. GR-15236, Penteli, Greece
\and
Royal Belgian Institute for Space Aeronomy, Avenue Circulaire 3, 1180 Uccle, Belgium
\and
Institut f\"ur Experimentelle und Angewandte Physik, Christian-Albrechts-Universit\"at zu Kiel, 24118 Kiel, Germany
\and
Observatoire de Paris, Meudon, France}

\abstract{We present a new catalogue of solar energetic particle events near the Earth, covering solar cycle 23 and the majority of solar cycle 24 (1996\textendash 2016), based on the 55\textendash 80 MeV proton intensity data gathered by the SOHO/ERNE experiment. In addition to ERNE proton and heavy ion observations, data from the ACE/EPAM (near-relativistic electrons), SOHO/EPHIN (relativistic electrons), SOHO/LASCO (coronal mass ejections, CMEs), and GOES soft X-ray experiments are also considered and the associations between the particle and CME/X-ray events deduced to obtain a better understanding of each event. A total of 176 SEP events have been identified as having occurred during the time period of interest; their onset and solar release times have been estimated using both velocity dispersion analysis (VDA) and time-shifting analysis (TSA) for protons, as well as TSA for near-relativistic electrons. Additionally, a brief statistical analysis has been performed on the VDA and TSA results, as well as the X-rays and CMEs associated with the proton/electron events, both to test the viability of the VDA and to investigate possible differences between the two solar cycles. We find, in confirmation of a number of previous studies, that VDA results for protons that yield an apparent path length of 1 AU $<$ $s \lesssim$ 3 AU seem to be useful, but those outside this range are probably unreliable, as evidenced by the anticorrelation between apparent path length and release time estimated from the X-ray activity. It also appears that even the first-arriving energetic protons apparently undergo significant pitch angle scattering in the interplanetary medium, with the resulting apparent path length being on average about twice the length of the spiral magnetic field. The analysis indicates an increase in high-energy SEP events originating from the far eastern solar hemisphere; for instance, such an event with a well-established associated GOES flare has so far occurred three times during cycle 24 but possibly not at all during cycle 23. The generally lower level of solar activity during cycle 24, as opposed to cycle 23, has probably caused a significant decrease in total ambient pressure in the interplanetary space, leading to a larger proportion of SEP-associated halo-type CMEs. Taken together, these observations point to a qualitative difference between the two solar cycles.}


   \maketitle
%%
%%________________________________________________________________

\section{Introduction}

Solar energetic particle (SEP) events, large injections of particles into interplanetary space from the Sun (see e.g. \citealt{{Reames1999}, {Reames2013}}), constitute an important component of space weather. They are closely associated with solar flare and coronal mass ejection (CME) activity, but the exact relationship between these and the particle acceleration processes remains an object of study. Aside from the valuable role that they have served in developing our understanding of the conditions and phenomena in the Sun and its atmosphere, as well as of particle transport and scattering in interplanetary space, SEPs also pose a threat in the form of tremendous increases in radiation dose rates. These may cause considerable harm to both manned and unmanned space missions, even rendering satellites unoperable. The crew and passengers of aircraft flying at high altitude or high geographic latitude may also be exposed to a non-negligible radiation hazard during the most energetic events (e.g. \citealt{Vainio2009}, \citealt{Reames2013}, \citealt{Mishev2014}, and references therein).

So far, the Sun has been remarkably quiescent during the current solar cycle, number 24 (e.g. \citealt{Richardson2013}, \citealt{Gopalswamy2015b}), which is here considered to have commenced in December 2008. This is in contrast to cycle 23 (1996\textendash 2008), which exhibited a great deal more of overall solar activity. A comparison between the two cycles in terms of SEP events and associated phenomena, such as X-ray flares and CMEs, is therefore called for, as it may add to our current knowledge regarding particle release and acceleration processes, as well as the interplanetary transport conditions.

The purpose of this paper is twofold. Firstly, it presents a catalogue of energetic (55\textendash 80 MeV) solar proton events for the years 1996 to 2016, supplemented with associated electron, electromagnetic and solar CME observations; secondly, these events are subjected to a statistical analysis, with focus on the methods for obtaining a solar release time for the energetic particles and on a rudimentary comparison of the two solar cycles. In both of these two regards, this article forms a follow-up study for \citet{Vainio2013}, and it is also intended to complement similar existing work done on comparing solar cycles 23 and 24 (e.g. \citealt{Chandra2013}, \citealt{Richardson2016}). We have extended the period of interest to cover the years 2011\textendash 2016 and also revisited the proton velocity dispersion analysis (VDA), electron event onset times, as well as electromagnetic (soft X-ray) and CME observations for the events listed in \citet{Vainio2013}. Based on data from the Energetic and Relativistic Nuclei and Electron experiment (ERNE; \citealt{Torsti1995}), the Large Angle and Spectrometric Coronagraph (LASCO; \citealt{Brueckner1995}), and the Electron Proton Helium Instrument (EPHIN; \citealt{Mueller-Mellin1995}) aboard the Solar and Heliospheric Observatory (SOHO) spacecraft, as well as the Electron, Proton, and Alpha Monitor (EPAM; \citealt{Gold1998}) aboard the Advanced Composition Explorer (ACE) and the X-ray flux measuring instrumentation on the Geostationary Operational Environmental Satellites (GOES) in active operation during the period of interest, our catalogue comprises a total of 176 energetic solar particle events between 1996 and 2016. This time span covers solar cycle 23 in its entirety and more than the first half of solar cycle 24, which allows us to gain some insight into the differences of these two cycles.

Particle release time analysis is performed for proton events using both VDA and time-shifting analysis (TSA) so as to obtain estimates for the particle release times near the Sun. We also investigate solar flare and CME activity for associations with the energetic particle releases and present some statistical comparisons and results between the previous and the on-going solar cycle, along with discussion.

Our goal has been to compile a comprehensive listing of solar proton events in the energy range of 55\textendash 80 MeV, which occurred during the aforementioned time period. The selection of this energy range was motivated by three principal reasons: 1) the significance of high-energy particles for space weather, 2) the relatively fast post-event decrease of intensities, enabling small, closely spaced events to be distinguished more easily than at low energies, and finally 3) the desire to maintain continuity with a number of previous ERNE-related studies.

It is important to note, however, that due to a number of gaps in the data coverage, as well as certain inherent limitations of the methodology applied to identifying and determining the events and their onset times, the event catalogue as presented herein is not entirely comprehensive. We nevertheless hope that it will both prove useful for further research and also stimulate it. In addition, we have been motivated in our work by the possibility that the apparent qualitative and quantitative changes from the previous solar cycle to the present one may lead to new discoveries and better understanding of the processes involved in SEP events.

The European Space Agency is establishing a Space Weather (SWE) Service Network\footnote{Session 2 -- SSA Space Weather Service Network, 13th European Space Weather Week meeting, 14--18 Nov. 2016, Oostende, Belgium} in the frame of its Space Situational Awareness (SSA) programme 
(\url{http://swe.ssa.esa.int/}). The goal of this incentive is to support end-users in a wide range of affected sectors to mitigate the effects of space weather on their systems, reduce costs and improve reliability. The network consists of five Expert Service Centres (ESCs): Solar Weather, Heliospheric Weather, Space Radiation, Ionospheric Weather, and Geomagnetic Conditions. The domain of the Space Radiation ESC (R-ESC; \url{http://swe.ssa.esa.int/space-radiation}) covers the monitoring, modelling, and forecasting of space particle radiation and micro-size particulates in the near-Earth space environment, as well as their effects on technological and biological systems. During the second period of the SWE segment of the SSA programme, new products are being provided to improve space weather services. The SEP catalogue described in this paper was recently integrated as a University of Turku (UTU) federated product under the umbrella of the R-ESC.

The structure of this paper is as follows. In Section \ref{catalog}, we  introduce the data used in this work and present the event catalogue. We describe our statistical analysis and its results in Section \ref{stats}, and outline our conclusions and outlook in Section \ref{conclusions}.

\section{The proton event catalogue for the years 1996--2016} \label{catalog}
\subsection{SOHO/ERNE proton data and event selection} \label{data_ev_selection}

The ERNE experiment (\citealt{Torsti1995}) aboard the SOHO spacecraft served as the primary instrument in this study. ERNE consists of two particle telescopes, the Low-Energy Detector (LED) and the High-Energy Detector (HED), which together are designed to cover the nominal energy range of some 1 MeV/nucleon to a few hundred MeV/nucleon for ions. For protons and helium ions, the upper limit of the energy range is about 140 MeV/nucleon. For the reasons explained in the previous chapter, the proton energy channel of 54.8\textendash 80.3 MeV (average energy 67.7 MeV) was chosen as the principal channel to be investigated.

To identify the SEP events that had occurred during the time of interest, we scanned visually through SOHO/ERNE intensity data collected during the period between May 1996 and December 2016. Our event selection criterion was such that the one-minute average intensity in the 54.8\textendash 80.3 MeV ERNE proton channel was required to surpass the quiet-time background of the relevant phase of the solar cycle by a factor of about three. The quiet-time background intensities\textemdash which are mostly the product of galactic cosmic rays\textemdash were estimated to range from $\sim$ 5 $\times$ 10$^{-4}$ cm$^{-2}$ s$^{-1}$ sr$^{-1}$ MeV$^{-1}$ in 1996\textendash 1997 near the solar minimum to $\sim$ 3 $\times$ 10$^{-4}$ cm$^{-2}$ s$^{-1}$ sr$^{-1}$ MeV$^{-1}$ in 2001\textendash 2003 near the solar maximum during solar cycle 23. For solar cycle 24, the corresponding values were $\sim$ 7 $\times$ 10$^{-4}$ cm$^{-2}$ s$^{-1}$ sr$^{-1}$ MeV$^{-1}$ in 2009\textendash 2010 and $\sim$ 3.5 $\times$ 10$^{-4}$ cm$^{-2}$ s$^{-1}$ sr$^{-1}$ MeV$^{-1}$ at the end of 2013.

The end times for the proton events were estimated, as well, for all energy channels. In our study, an SEP event is defined as having ended when the intensity of protons in the 12.6\textendash 13.8 MeV (average 13.3 MeV) energy channel falls below twice that of the pre-event background\footnote{This was derived by calculating the mean intensity for the 2 hours immediately preceding the event onset. In cases where the data were unavailable or indicated that the previous event had not yet ended, estimated quiet-time background intensity, calculated separately for each semiannual period, was substituted for the pre-event background.} of this energy channel for the first time. Observing the onset and end in different energy channels is justified by the fact that in a typical event, after the most energetic particles have already arrived at and passed the orbit of the Earth, considerable numbers of slower, less energetic particles are still on the way. To consider only the greater energy channel would mean to exclude these slower protons which nevertheless contribute significantly to the total ion fluence and energy release of the event and may potentially cause harm to spacecraft and exposed human beings, such as astronauts. If another high-energy proton event occurs before the previous one has ended according to the criterion explained above, the onset of the next event is taken as the end of the preceding one.

It is important to note that the data collection or transmission from the SOHO spacecraft to the Earth was interrupted on several occasions, most notably between the late June and early October 1998, as well as late December 1998 and early part of February 1999. Any SEP events during such periods will have gone unrecorded, as well as some very minor events in the immediate aftermath of extremely large ones that have caused HED to saturate. However, as the saturation can only occur at the height of large events, its overall effect on event selection and onset timing is negligible. A comprehensive listing of ERNE data gaps is given in Table \ref{ERNE_datagaps}.

% The ERNE data gap table:

\begin{table*}
\caption{Major (longer than $\sim$ 18 h) continuous ERNE data gaps. All times are UT.}             
\label{ERNE_datagaps}      
%\centering
\footnotesize        
\begin{tabular*}{\textwidth}{l @{\extracolsep{\fill}} llll}
\hline\hline       

\hline
28\textendash 31-May-1996&28\textendash 29-Jul-2001&26\textendash 27-Sep-2010&14\textendash 15-Oct-2013\\
03-Jun-1996&10\textendash 17-Aug-2001&10\textendash 15-Dec-2010&29-Oct\textendash 04-Dec-2013\tablefootmark{e} \\
15\textendash 16-Feb-1997&01\textendash 03-Jan-2002&09\textendash 13-Jan-2011&29\textendash 31-Jan-2014\\
25-Feb\textendash 04-Mar-1997&24\textendash 25-Jan-2002&13\textendash 14-Jan-2011&21-Mar-2014\\
25\textendash 28-Oct-1997&05\textendash 12-Feb-2002&07-Apr-2011&21-Sep-2014\\
19\textendash 21-Nov-1997&24-May-2002&17-Jul-2011&02\textendash 03-Nov-2014\\
27-Nov-1997&01\textendash 02-Jul-2003&21-Jul-2011&17\textendash 19-Dec-2014\\
11\textendash 12-Dec-1997&14\textendash 15-Jan-2003&30-Nov-2011\textendash 05-Jan-2012\tablefootmark{b} &09\textendash 15-Jan-2015\\
04\textendash 05-Feb-1998&28-Feb\textendash 11-Mar-2003&22\textendash 24-Jan-2012&24\textendash 26-Jan-2015\\
25\textendash 28-Feb-1998&05-Jul-2003&25\textendash 27-Jan-2012&12\textendash 14-Feb-2015\tablefootmark{a}\\
24-Jun\textendash 09-Oct-1998&09\textendash 10-Jul-2003&28-Jan\textendash 10-Feb-2012\tablefootmark{c} &14-Feb-2015\\
14\textendash 21-Nov-1998&22\textendash 24-Oct-2003&15\textendash 16-Feb-2012&23\textendash 24-Mar-2015\\
01\textendash 03-Dec-1998&19\textendash 22-Jan-2004&04-May-2012&05-Jun-2015\\
21-Dec-1998\textendash 08-Feb-1999 \,&26-Mar-2004&07-May-2012&04, 05-Sep-2015\\
14\textendash 18-Feb-1999&02-Apr-2004&12\textendash 14-May-2012&10-Sep-2015\\
21\textendash 26-May-1999&22\textendash 29-Apr-2004&22-May-2012&02-Oct-2015\\
03\textendash 04-Jun-1999&06\textendash 12-Aug-2004&07\textendash 08-Jul-2012&06\textendash 09-Nov-2015\\
19\textendash 20-Aug-1999&08\textendash 09-Dec-2004&02\textendash 07-Nov-2012&14\textendash 16-Nov-2015\\
15\textendash 20-Nov-1999&22\textendash 23-Dec-2004&06\textendash 07-Dec-2012&30-Nov\textendash 03-Dec-2015\tablefootmark{a} \\
28\textendash 29-Nov-1999&31-Jul\textendash 03-Aug-2005&09-Dec-2012\textendash 31-Jan-2013\tablefootmark{d} &07-Dec-2015\\
01\textendash 02-Dec-1999&08\textendash 09-Mar-2006&01\textendash 05-Feb-2013&10\textendash 11-Dec-2015\\
07\textendash 08-Jan-2000&13\textendash 14-Aug-2006&06\textendash 07-May-2013&05\textendash 08-Jan-2016\\
22\textendash 25-Feb-2000&31-Aug-2006&10\textendash 13-May-2013&30\textendash 31-Mar-2016\\
13\textendash 31-Mar-2000&27\textendash 29-Jan-2007&08\textendash 10-Jun-2013&22\textendash 25-Jul-2016\\
18\textendash 19-Apr-2000&11-Nov-2008&17\textendash 22-Jul-2013&06\textendash 07-Oct-2016\\
22\textendash 24-May-2000&20-Feb-2009&13\textendash 14-Aug-2013&18-Oct-2016\\
14\textendash 15-Jan-2001&07-Sep-2006&07\textendash 09-Sep-2013&21\textendash 22-Dec-2016\\
02\textendash 03-Jul-2001&12\textendash 15-Sep-2006\tablefootmark{a}&29\textendash 30-Sep-2013& \\
\hline             
\end{tabular*}

\normalsize
\noindent \tablefoottext{a}{Some data are available during this period.}\\
\tablefoottext{b}{Data available for less than 20 minutes on 07-Dec-2011.}\\
\tablefoottext{c}{Data available for a few minutes on 30-Jan-2012.}\\
\tablefoottext{d}{Some 12 minutes of data available on 11-Dec-2012.}\\
\tablefoottext{e}{Data available for a few minutes on 30-Oct-2013.}\\
\end{table*}

\subsection{Onset, solar release, and event end time determination}
The proton event onset times as observed by SOHO/ERNE were determined using the so-called Poisson-CUSUM method. This essentially measures the statistical quality of a process to decide whether or not the process is under control and determines the moment of time of a failure occurring. Here, the pre-event background level proton flux is analogous to a controlled process, while a flux increase at event onset corresponds to a departure from control. The onset determination algorithm and the criteria used have been previously described in detail in \citet{Huttunen-Heikinmaa2005}.

When the observed event onset time is known, the velocity dispersion analysis  for a given particle species and kinetic energy $E$ can be performed. The observed onset time at 1 AU can be written as

\begin{equation}
t_{\rm onset} (E) = t_0 + 8.33 \, {\rm[min/AU]} \, s \, \beta^{-1}(E),
\end{equation}
where $t_\textrm{0}$ is the particle release time (in minutes) from the acceleration site, $s$ is the apparent path length (in AU) travelled by the particles and $\beta^\textrm{-1}(E)$ is their reciprocal speed in units of $c^{-1}$. Linear fitting of the observed onset times as a function of the inverse speed thus yields an estimate for both $t_\textrm{0}$ and $s$. The basic assumptions of VDA include a simultaneous release of particles of all  energies and the same apparent path length for all particles. The latter assumption requires that the first-arriving particles either undergo no scattering, or that the scattering affects these particles independently of their kinetic energy.

The VDA was performed using 20 proton energy channels, spanning the range 1.58 MeV to 131 MeV (see Table \ref{VDA_channels}). Of this, LED covers the energies between 1.58 MeV and 12.7 MeV, while HED covers those between 13.8 MeV and 131 MeV, both providing ten individual channels. The time resolution was one minute for all channels.

\begin{table}
\caption[]{ERNE energy channels used for proton VDA.}
\label{VDA_channels}
\centering                          % used for centering table
\begin{tabular}{l c c c c}        % centered columns (5 columns)
\hline\hline                 % inserts double horizontal lines
& Channel & Energy range [MeV] & Average energy [MeV] & Reciprocal speed [$c^\textrm{-1}$] \\    % table heading 
\hline                        % inserts single horizontal line
LED & 1 & 1.58\textendash 1.78 & 1.68 & 16.7 \\
& 2 & 1.78\textendash 2.16 & 1.97 & 15.5\\
& 3 & 2.16\textendash 2.66 & 2.41 & 14.0\\
& 4 & 2.66\textendash 3.29 & 2.98 & 12.6\\
& 5 & 3.29\textendash 4.10 & 3.70 & 11.3\\
& 6 & 4.10\textendash 5.12 & 4.71 & 10.0\\
& 7 & 5.12\textendash 6.42 & 5.72 & 9.10\\
& 8 & 6.42\textendash 8.06 & 7.15 & 8.15\\
& 9 & 8.06\textendash 10.1 & 9.09 & 7.24\\
& 10 & 10.1\textendash 12.7 & 11.4 & 6.47 \\
\hline
\noalign{\smallskip}
HED & 11 & 13.8\textendash 16.9 & 15.4 & 5.59\\
& 12 & 16.9\textendash 22.4 & 18.9 & 5.06\\
& 13 & 20.8\textendash 28.0 & 23.3 & 4.57\\
& 14 & 25.9\textendash 32.2 & 29.1 & 4.11\\
& 15 & 32.2\textendash 40.5 & 36.4 & 3.69\\
& 16 & 40.5\textendash 53.5 & 45.6 & 3.32\\
& 17 & 50.8\textendash 67.3 & 57.4 & 2.99\\
& 18 & 63.8\textendash 80.2 & 72.0 & 2.70\\
& 19 & 80.2\textendash 101 & 90.5 & 2.44\\
& 20 & 101\textendash 131 & 108 & 2.26 \\
\hline
\end{tabular}

\end{table}

In the majority of cases, one or more data points were discarded from the VDA fitting. This was mainly due to an elevated background from a previous event, rendering no additional enhancement detectable at low energies, or the fact that no noticeable intensity enhancement at the highest energies was observed. A reasonable velocity dispersion relation could not be derived for all events, and in some cases, a statistically acceptable fit exists but the resulting apparent path length is unphysical, namely less or very much more than 1 AU. However, as the continuing evaluation of the performance of VDA is one of the objectives of this work, these cases have not been \textit{a priori} excluded from the listing or the statistical analysis.

The time-shifting analysis (TSA) for protons and electrons was performed by back-shifting the previously determined onset (near the Earth) to the vicinity of the Sun. The release time for a particle species with kinetic energy $E$ can be expressed as follows:

\begin{equation}
t_{\rm rel}(E) = t_{\rm onset}(E) - 8.33 \, {\rm[min/AU]} \, L \, \beta^{-1} (E).
\end{equation}
Here, $\beta^\textrm{-1}$ is again the reciprocal speed of the particles, and $L$ is the length of the magnetic field line connecting the source and the observer, computed from the detected speed of the solar wind $u_{\rm SW}$ during the event as

\begin{equation}
L(u_{\rm SW}) = z(r_{\rm SC}) - z(r_{\odot}),
\end{equation}
where $z(r)$ is the distance along the Archimedean spiral from the centre of the Sun, $r_{\rm SC}$ is the radial distance of the observing spacecraft from the Sun, and $r_{\odot}$ is the solar radius (taken to be 6.957 $\times$ 10$^5$ km). $r_{\rm SC}$ was approximated by calculating the Earth\textendash Sun (centre-to-centre) distance and subtracting from this the Sun-directional distance of the spacecraft from the centre of the Earth.

The solar wind speed, relevant to the calculation of $z(r)$, was determined using data from the Solar Wind Electron, Proton and Alpha Monitor (SWEPAM) aboard ACE (\citealt{McComas1998}; \url{http://www.srl.caltech.edu/ACE/ASC/level2/lvl2DATA_SWEPAM.html}), with Wind/SWE data (\citealt{Ogilvie1995}; \url{http://omniweb.gsfc.nasa.gov/ftpbrowser/wind_swe_2m.html}) substituted for the events for which the former were not available. In each case the data were averaged over a 12-hour period, centered on $t_{\rm onset}$ for 54.8\textendash 80.3 MeV protons observed by ERNE; in cases where the precise $t_{\rm onset}$ for protons was unknown but known for electrons, the UT calendar day (24 hours) during which the onset had most likely occurred was used to obtain $L$ for electron TSA. The Earth\textendash Sun and Earth\textendash spacecraft distances were averaged over the period of interest in the same manner.

In addition to proton data, we examined electron intensity data recorded by both ACE/EPAM (near-relativistic electrons) and SOHO/EPHIN (relativistic electrons). These were analysed for onset times, and ACE/EPAM data additionally for maximum intensities during the events as well as particle release times, so as to form a more complete picture of each event than would be possible by relying on proton data alone.

By default, the ACE/EPAM data (\citealt{Gold1998}) used in our analysis were sectored one-minute intensities observed by the LEFS60\footnote{The 
numbers in the telescope names denote the angles (in degrees) of the boresight directions of the telescopes measured from the spin axis of the 
spacecraft.} (Low-Energy Foil Spectrometer) telescope. These data were checked for ion contamination by comparing them visually with proton 
intensities recorded by the LEMS120 (Low-Energy Magnetic Spectrometer) telescope, as well as with electron intensities recorded by DE30 (Deflected 
Electrons). If any indications of possible ion contamination were detected in the LEFS60 intensity data, DE30 data were used instead. While ion 
contamination occasionally presents an issue for electron data recorded by LEFS60, it has the advantage of having a larger geometric factor than DE30 ($\sim$ 0.40 cm$^2$ sr for electrons for the former, compared to $\sim$ 0.14 cm$^2$ sr for the latter) and was for this reason preferred in our study. A similar electron intensity profile comparison to determine possible ion contamination is described in \citet{Malandraki_etal_2000}. The authors performed their comparison on data recorded during an SEP event by the Ulysses/HI-SCALE instrument which is virtually identical to ACE/EPAM (the latter being the flight spare of Ulysses/HI-SCALE).
 
The electron event onset times for ACE/EPAM were determined by first estimating the yearly background intensity ($I_{\rm bg}$), calculating the sliding time-average intensity $I_{\rm av}(t)$ over 60 minutes, and then comparing $I_{\rm av}(t)$ and the corresponding standard deviation to data points following the time window by another 60 minutes. When five consecutive data points all showed an intensity equalling or exceeding $I_{\rm av}(t) + n \sigma$, where $n$ was a user-defined constant (with the default value of 2.0), the time stamp of the first point was defined as the preliminary onset of the event. Gathering the intensity data points covering the range [61,60+$d$] minutes before the preliminary onset\textemdash where $d$ was a user-defined parameter, here set to either 180 or 240\textemdash and subtracting $I_{\rm bg}$ from these, an exponential curve of the form $I'(t) = \exp(A-Bt)$ was fitted with the least squares method to the intensity data. $I'(t)$ was then subtracted from the intensity data so as to remove trends, yielding detrended intensity $I_{\rm detrend}(t) = I(t) - I_{\rm bg} - I'(t)$. Finally, the onset search was performed using $I_{\rm detrend}(t)$ in a similar manner as that described above, with the exception that the comparison time window was kept fixed at 61\textendash 120 minutes before the preliminary onset. The definitive onset was taken to be the time stamp of the first of five consecutive data points equal to or exceeding the 60-minute pre-onset quantity $I_{\rm detrend}(t) + n \sigma$.

The principal ACE/EPAM energy channel used for determining the onsets was 0.18\textendash 0.31 MeV. However, in five cases (events 1, 10, 41, 118, and 141) it was deemed necessary to resort to the 0.10\textendash 0.18 MeV energy channel so as to obtain a clear result, due to an elevated pre-event background or weak enhancement of the intensity in the higher energy channel. The energy ranges correspond to mean electron energies of 0.23 MeV and 0.13 MeV and thus mean speeds of $0.73 \, c$ and $0.61 \, c$, respectively. These speeds are so high, and the number of available energy channels (four) so small, that attempting to perform VDA for ACE/EPAM electron data was not expected to produce reliable results in most cases. Instead, the onset times were determined for each event and, after discarding cases where no reasonable result for $t_{\rm onset}$ could be obtained, only TSA was performed.

Near-relativistic electron event information derived from ACE/EPAM observations is complemented by event onset times determined from SOHO/EPHIN electron data (\citealt{Mueller-Mellin1995}). The energy channel of interest was 0.7--3.0 MeV. The onset was defined in a similar, but slightly simpler, manner as for ACE/EPAM data: the average intensity over a specified time window $I_{\rm av}$ and the corresponding standard deviation $\sigma$ multiplied by a user-defined constant $n$ were compared to immediately following data points, and the time stamp of the first data point to exceed the intensity value of $I_{\rm av} + n \sigma$ (where in this case $n$ = 4.0) was taken as the time of onset. No detrending or release time analysis was performed for these electron observations, but they are included in this article to continue the analysis of the datasets used in \citet{Vainio2013}.

The durations of the SEP events were estimated according to the basic criteria explained in \ref{data_ev_selection} in the following manner. The preliminary event end times were derived by locating the time point where the proton intensity in the 12.6\textendash 13.8 MeV energy channel first fell below the pre-defined threshold value (2.0 $\times$ the pre-event average background intensity or the quiet time background intensity, if the former was substantially elevated) after the 55\textendash 80 MeV proton event onset. The period between the onset and the preliminary event end was then scanned visually for data gaps and new low-energy SEP events not visible in high energies. If any were found, the event ending time was shifted to the nearest noon or midnight UT before the next low-energy event or to the beginning of an extensive\textemdash i.e. longer than about an hour\textemdash data gap, if it was considered possible that significant new proton activity might have occurred during the gap. In cases where another 55\textendash 80 MeV proton event occurred before any of these conditions was met, the previous event was regarded as ending at the onset of the next.

\subsection{Ion fluences and iron-to-oxygen ratios}
Fluences (i.e., differential particle intensities integrated over a given range of energy and time, expressed here in cm$^\textrm{-2}$ sr$^\textrm{-1}$) of energetic protons and oxygen ions, as well as iron/oxygen ratios, were calculated for each event using ERNE data. For protons, the energy range of interest was from $\sim$ 10 MeV to 140 MeV (prior to 19 April 2000) or 131 MeV (after 19 April 2000); the two different upper limits follow from a modification to the ERNE onboard software with a data format change on the date given above. As the bulk of the ion fluence is expected to occur at low energies, however, this difference is unlikely to have a large impact on the results. The estimated quiet-time background was first subtracted from the intensities of each energy channel. Short time intervals with known issues with the ERNE instrument were treated as data gaps, and these, together with actual (non-critical) data gaps, were compensated by logarithmic interpolation using neighbouring "good" data points. Because of the limitations of this method, it was not applied to events during which substantial loss of coverage at the event onset or intensity maximum had occurred. Due to the large uncertainties involved, no fluence estimate is given for these events. When applicable, the proton fluence results were checked against those given in \citet{Papaioannou2016} (based on GOES proton data), and when they differed by more than a factor of about two, the event in question was subjected to further analysis to determine whether the obtained result was indeed reliable. Any results thus deemed unreliable, usually because of ERNE saturation during the event, are marked to that effect in the catalogue.

The energy range of interest for oxygen and iron ions was 5\textendash 15 MeV/nucleon. Since the intensities of heavy elements are typically several orders of magnitude smaller than those of protons, the interpolation method used with proton intensities could not be employed here as such. Instead, we first calculated $^4$He fluences in the energy range of 5\textendash 17 MeV/nucleon (this being the closest equivalent available to the heavy ion energy range mentioned above) in the same manner as proton fluences and then used the ratio between the measured and corrected helium fluences as a correction factor for the measured oxygen fluences. Helium was used as a proxy because it has a similar charge-to-mass ratio to oxygen ion (or the same, assuming both are fully ionized). Finally, the iron-to-oxygen ratios were calculated from the measured event-averaged intensities for these particle species. In cases where no counts for oxygen were measured for the duration of an entire event, the value given in the catalogue for oxgyen fluence is the oxygen one-count upper limit, and the Fe/O ratio is omitted. Conversely, when oxygen counts were measured during an event but iron counts were not, we calculated an upper limit for the Fe/O ratio using one-count upper limit for iron intensity.\footnote{Even though neither oxygen fluences nor the Fe/O ratios are discussed or analysed further in this work, they are of interest here since they are prominently featured in the online version of our SEP event catalogue and serve as the main topic of an upcoming study.}

\subsection{Soft X-ray and CME data}
\label{x-ray_cme}
The tabulated information regarding X-ray flare events is based on GOES satellite data made available by the United States National Oceanic and Atmospheric Administration (NOAA; \url{http://www.ngdc.noaa.gov/stp/space-weather/solar-data/solar-features/solar-flares}), with the flare/SEP event association deduced by us. The electromagnetic (EM) wavelength band of interest was 0.1\textendash 0.8 nm, i.e. soft X-rays. However, only flares with NOAA/GOES magnitude classification of C1.0 or greater (peak flux at least 10$^\textrm{-6} \, {\rm W/m^\textrm{2}}$) were taken into consideration. Flare onset, classification, and location were in the majority of cases provided directly by NOAA listings, and the time derivative maximum of the soft X-ray flux was calculated based on the flux data smoothed with five-minute sliding average for use in the particle injection time estimation. In cases where the flare classification and onset information were available but the coordinates unavailable in the NOAA X-ray or H-alpha listings\footnote{Events 115, 116, 117, 127, 129, 139, 142, 151, 155, 156, 160, 168, 172, 174, and 175.}, the derived position provided in the SolarSoft Latest Events Archive (\url{http://www.lmsal.com/solarsoft/latest_events_archive.html}) was given as the location of the flare. If the flare that was most likely associated with an SEP event preceded the observed proton onset by more than six hours, the event was marked with the note "t$_p$-t$_x$ $>$ 6 h" in the tables and the flare omitted from any detailed statistical considerations. These results, while derived independently, were checked against the event information provided in \citet{Cane2010} and \citet{Richardson2014}.

The SEP event-related flares that occurred at the western solar limb or near enough behind it (western longitude between 90 degrees and $\sim$ 120 degrees) to have a NOAA classification and onset time available are marked as "long. $\geq$ 90" in the catalogue. These cases, including their location, were also confirmed using the listings in \citet{Cane2010} and \citet{Richardson2014}. Lastly, SEP events with no clearly identifiable GOES flare carry the note "No GOES flare"; note, however, that in most (if not all) cases there is farside flare activity involved, according to the articles mentioned above.

The time derivative of the X-ray intensity provides a tool for studying electron acceleration in the solar atmosphere. In solar observations, there is a well-known correlation between hard X-ray and microwave intensity and the time derivative of the soft X-ray intensity during the impulsive phase of an X-ray event. Pointed out by \citet{Neupert1968} and since expounded by several authors, this correlation is commonly called the Neupert effect and is thought to arise when non-thermal accelerated electrons lose energy via bremsstrahlung upon encountering dense surrounding plasma, rapidly heating it (see e.g.\citealt{Veronig2002} for details). Thus, we may reasonably assume that the peak of acceleration of near-relativistic electrons at the Sun occurs approximately when the time derivative of the soft X-ray intensity reaches its local maximum. Provided that open magnetic field lines are present at the acceleration site, this corresponds to the time of release of electrons into interplanetary space.

The SOHO LASCO CME Catalog (\url{http://cdaw.gsfc.nasa.gov/CME_list/}) was used for identifying and listing the features (first appearance, estimated linear speed, width, and central position angle) of the CMEs likely associated with each SEP event. After an estimate of the approximate time of particle injection had been derived using VDA and TSA results and refined using flare and radio frequency data, the CME Catalog was searched for relevant CME activity. If the listing included a CME with an estimated solar surface departure time differing from the radio and flare activity by less than $\sim$ one hour, it was analysed further and then selected for our catalogue, if its association with the SEP event appeared reasonably certain in light of other evidence (see \ref{ev_case_study} for an example).

We list the first detection of the CME as observed on the SOHO/LASCO C2 coronagraph, with the time figure rounded to the nearest minute. In a few cases, the CME was not observed on C2; for these events, the first observation is recorded for the C3 (see notes for Table 4). For a detailed introduction to the CME Catalog and discussion, see \citet{Gopalswamy_etal_2009}.

While radio frequency observations are not included in the tables presented in this article, spectral plots of Wind/WAVES data\footnote{These are available at \url{http://cdaweb.gsfc.nasa.gov}; on the main page, select "Wind" and "Radio and Plasma Waves (space)," then "Wind Radio/Plasma Wave, (WAVES) Hi-Res Parameters."} were scanned visually during periods of SEP, CME, and flare X-ray activity to find out whether or not type III microwave bursts were detected close in time to both the onset of the candidate flare and the surface lift-off of the candidate CME. Such radio bursts are in fact a general characteristic of $\gtrsim$ 20 MeV proton events (\citealt{Cane2002}). The presence or absence of type III burst activity was used to determine whether or not there was a plausible association between the candidate flare, candidate CME, and the SEP event itself.

\subsection{Event analysis case study: event 3}
\label{ev_case_study}
In the following, we briefly describe the analysis performed on a typical SEP event and the derivation of the tabulated quantities. The event concerned is event 3, which occurred on 6 November 1997. All times mentioned below and in the rest of this work are UT, and all particle release time estimates include the approximate light travel time, 500 seconds.

A visual scan of the one-minute ERNE 55\textendash 80 MeV proton intensity data identified an enhancement of about two orders of magnitude near noon on 6 November 1997 as a candidate event. After it was subjected to closer study, the Poisson-CUSUM method yielded the onset time of 12:37 in this energy channel, as well as the corresponding onset time values for the ERNE VDA channels, which allowed VDA to be performed. The resulting apparent path length is 2.61$\pm$0.24 AU and the estimated particle release time of 12:21$\pm$00:17. On the other hand, TSA yields a magnetic field line length of 1.20 AU and estimated particle release time of 12:18.

For electrons, the onset and release time results are remarkably similar. The onset of ACE/EPAM 0.18\textendash 0.31 MeV electrons occurs at 12:24, and that of SOHO/EPHIN 0.7\textendash 3.0 MeV electrons at 12:23. For the former, TSA indicates a particle release time of 12:18. Thus, the available information for both protons and electrons suggests a particle release at or very near 12:20.

Turning to the SOHO LASCO CME Catalog, we next searched the time period near noon, 6 November 1997, for CME activity. An immediately obvious candidate for the SEP event-associated CME is a fast halo event first detected by LASCO at 12:11; in fact, no other CME is listed as having been detected for several hours either before or after this one. A linear fit to observation data points presented in the SOHO LASCO CME Catalog gives the speed of the CME as 1556 km/s and the departure time from the solar surface as 11:30. This is corroborated by a visual scan of Wind/WAVES data, which indicates considerable type III radio burst activity beginning at circa 11:52.

NOAA listings show that during a few hours within noon of 6 November 1997, two soft X-ray flares with NOAA classification greater than C1.0 were detected by GOES: a C4.7 class flare, beginning at 11:31 and ending at 11:44; and an X9.4 class flare, beginning at 11:49 and ending at 12:01. The location of the X9.4 class flare is given in the listing as 18S 63W, and our analysis indicates the maximum of the time derivative of the soft X-ray flux as having occurred at 11:53. The larger and later of the two flares coincides with the estimated departure from the solar surface of the CME, as well as the derived particle release times for both 55\textendash 80 MeV protons and 0.18\textendash 0.31 MeV electrons. Therefore, combining the results mentioned above, it seems rather safe to conclude that the SEP event, the CME, the radio activity, and the X class flare are very likely associated with one another. In addition, it is noted that the SEP event information given in \citet{Cane2010} is in agreement with this interpretation.

Finally, the end of this proton event is estimated to have occurred at 22:26 on 13 November 1997. While ERNE proton intensity data are available for only 87.3 \% of the time between event onset and end, none of the data gaps are fortunately located at the critical rise and peak stages of the event, nor are there long continuous periods of missing data. Thus, estimates for the proton and oxygen fluence and the iron/oxygen ratio could be calculated with satisfactory certainty in this case.

\begin{figure}
\centering
\includegraphics[width=0.7\columnwidth]{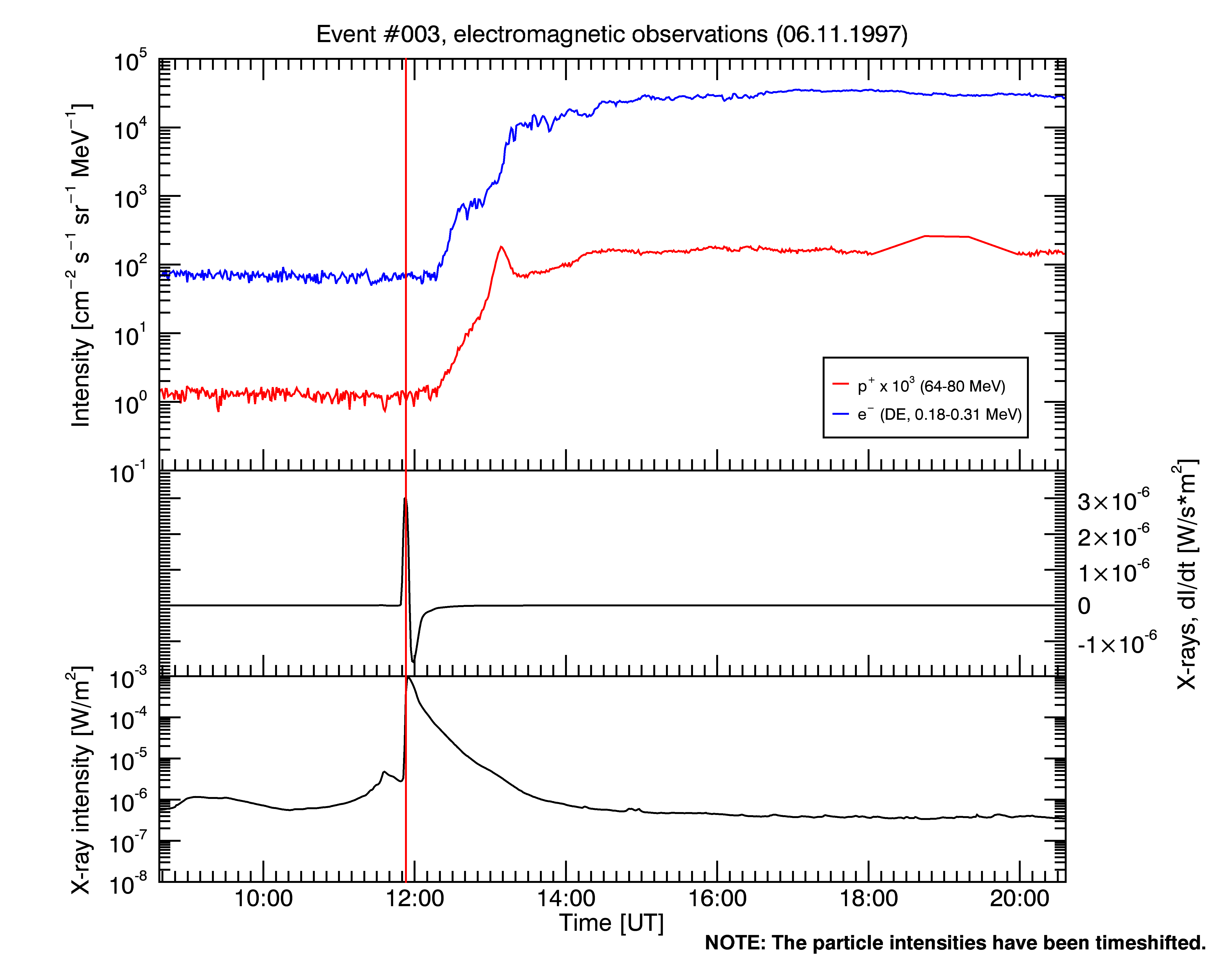}
\caption{\small The particle and X-ray observations related to event 3 (6 November 1997). Shown here are proton and electron intensities (top panel, the former multiplied by 10$^3$), the time derivative of the soft X-ray intensity (middle panel), and the soft X-ray intensity (bottom panel). The moment of the maximum of d$I$/d$t$ for X-ray intensity is marked with a vertical red line. The particle intensities have been backshifted in time by ($L$/$v$ - 500) seconds, where $L$ = 1.20 AU and $v$ = the mean speed of the particle species of interest.}
\label{ev3_fig}
\end{figure}

An overview of event 3 is presented in Figure \ref{ev3_fig}. It shows backshifted proton and electron intensities juxtaposed with X-ray observations; the time derivative maximum of the soft X-ray intensity is marked with a red line.

\subsection{The event catalogue}
\label{ev_cat_2_6}
The proton event catalogue is presented in Table 3 and Table 4, provided as additional online material. Table 3 lists the proton events, organized by date, and provides the reference number (ID), date and time of detected 55\textendash 80 MeV proton onset, approximate maximum intensity $I_{\rm max}$ (in units of pfu MeV$^\textrm{-1}$, where pfu = cm$^\textrm{-2}$ s$^\textrm{-1}$ sr$^\textrm{-1}$), and event end date and time; the spiral field line length $L$ (in AU) and the TSA-derived particle release time $t_{\rm rel}$; the apparent path length $s$ (in AU), release time $t_{\rm 0}$ together with its standard error, and the square of the sample correlation coefficient $R^2$ derived from VDA for 55\textendash 80 MeV protons; ERNE proton and oxygen fluences for $>$ 10 MeV and 5\textendash 15 MeV/nucleon (in cm$^\textrm{-2}$ sr$^\textrm{-1}$), respectively; and iron-to-oxygen ratios for 5\textendash 15 MeV/nucleon are also given here. Table 4 lists the corresponding electron events together with flare and CME information. It shows the event reference number (ID), electron event onset date and time, estimated maximum electron intensity $I_{\rm max}$, and TSA-derived release time $t_{\rm 0}$ for 0.18\textendash 0.31 MeV electrons; the onset time for 0.7\textendash 3.0 MeV electrons; the time of maximum slope of the soft X-ray intensity (Max. d$I$/d$t$), NOAA magnitude classification, and solar location of the event-related GOES flare; and the first observation time, speed $v$ (in km/s), and width and position angle (in degrees) of the event-associated CME. All particle release times given in the tables include an additional 500 seconds to account for the light travel time to the observer.

For six proton events, the accompanying electron event could not be identified with satisfactory certainty, mainly due to the overall weakness and slow rise of the electron intensity enhancement. In three cases (events 119, 132, and 144), the onset criterion for ACE/EPAM electron intensity was triggered some seven hours or more before the proton event onset, suggesting that the electrons probably originated from a different injection; electron onset times for these entries were omitted from Table 4. In one case (event 149, 30 September 2013), the onset of the proton event was unknown owing to a data gap, rendering further analysis of the electron and X-ray/CME observations (aside from a rough estimation of the maximum intensity of electrons) impossible. Also note that for two events (numbers 71 and 91, 17 April 2002 and 20 November 2003, respectively), the first near-relativistic electrons to arrive at ACE may be from injections that precede the main proton event. A similar situation, concerning relativistic electrons, may have occurred for SOHO/EPHIN in events 41, 65, 73, 91, 94, 127, and 136. In each of these cases, there appear two or more successive electron intensity rises, but the intensity remains elevated to such an extent for several hours that the onset determination algorithms for the respective instruments are unable to distinguish between the apparent events. Thus, for these catalogue entries, the electron onset time given refers to the earliest possible moment of actual event onset and is prefixed with "$\geq$" in Table 4.

Altogether there were 34 cases in which no visible disk or over-the-limb GOES flare could be associated with the SEP event with confidence; in another eight cases the likely candidate flare had occurred more than six hours prior to the proton event onset and was for this reason not listed. In addition, four cases could not be investigated due to missing data, and two (events 56 and 66, discussed further below) remained uncertain in this respect due to  very long delays between particle release and event onset near the Earth. This leaves 128 events for which at least some information of the associated GOES flare is available and the flare identification is relatively secure. Thirteen of these flares certainly or very likely occurred at or immediately behind the solar limb; thus, full flare information (time of the maximum of the intensity time derivative, classification, location) is listed for 115 events (72 during 1996\textendash 2008, 43 events during 2009\textendash 2016). Independent longitude estimation was not attempted for the farside flares.

The observation that SEP events and CMEs are associated and that the peak intensities of the events are correlated with CME speed is well established (see e.g. \citealt{Kahler1984}, \citealt{Kahler2001}). It is therefore unsurprising that in our data set, virtually every SEP event that extended to proton energies above 55 MeV was accompanied by an identifiable CME. In events 56 (10 August 2001) and 66 (10 January 2002), the proton intensity increase was extremely slow and prolonged in high energies, and the particle injection that caused these events had possibly occurred on the previous calendar day, if not before. As we could not confirm the timing and indeed the identity of the relevant particle injection to our satisfaction, no CME or flare is listed for these events. We note, however, that \citet{Cane2010} suggest a slow partial halo CME (speed 479 km/s) first observed at 10:30 on 9 August 2001 and a possible farside flare for event 55, and a CME (speed 1794 km/s) first observed at 17:54 on 8 January 2002 together with an M1 class flare over the eastern solar limb for event 66. In contrast, \citet{Papaioannou2016} list event 66 as being associated with a C7 class GOES flare occurring at 9:27 on 10 January 2002 but do not offer a suggestion for the accompanying CME. 

In addition to these two outstanding cases, SOHO/LASCO data are missing for five events, and there are otherwise insufficient data to make a positive determination as to the particle event-associated CME in three cases. All the other 166 events exhibit unambiguous CME activity. Of these, 136 (91 in 1996\textendash 2008, 45 in 2009\textendash 2016) were also known to be associated with a GOES flare on the visible solar disk or over the western solar limb (the cases of missing X-ray intensity data are excluded from these figures). Estimates for both the speed and apparent angular size are available for all CMEs listed in our catalogue, aside from seven in  1996\textendash 2008, for which the SOHO LASCO Catalog only reports a lower limit for the size, in addition to the estimated speed.

\section{Statistical results and discussion}
\label{stats}

\subsection{VDA results}
We investigated 114 events of solar cycle 23 and 62 events of solar cycle 24 (until the end of the year 2016), a total of 176 events. A VDA result was found for all but three events of cycle 23 and 12 events of cycle 24, leaving  161 events (111 for cycle 23, 50 for cycle 24) available for analysis. The statistical considerations presented in this subsection are similar in nature and given in approximately the same order as in \citet{Vainio2013} so as to facilitate comparison of results and show progress made since the publication of that paper.

The algorithm used for onset time determination may completely fail in some circumstances. Such circumstances are, e.g., high pre-event background, very slow rise of the intensities, large sudden fluctuations in the intensities just before the onset, and low overall fluxes (insufficient sensitivity of the instrument). Unphysical results may be caused by one or a few onset times given by the algorithm, which are clearly erroneous. Removing these data points from the VDA may give physically more significant results, but of course also makes these particular results more subjective in the sense that one has to decide which data points to remove from the analysis.

The cases with no VDA result were omitted from consideration. Additionally, there were nine cases of very long ($>$ 5 AU) and seven cases of very short ($<$ 1 AU) apparent path lengths among the 161 events. In four of the former and two of the latter events, at least three energy channels were discarded from the VDA, and in three cases the proton intensity measured by ERNE increased very slowly at event onset. However, it is noteworthy that redoing the VDA with a more critical view on the goodness of the data points has decreased the number of the unphysically short apparent path lengths: in \citet{Vainio2013}, a path length of less than 1 AU is reported to have resulted in eight of 111 events.

\begin{figure}
%\begin{minipage}{\textwidth}
\centering
\begin{tabular}{c}
\includegraphics[width=0.6\columnwidth]{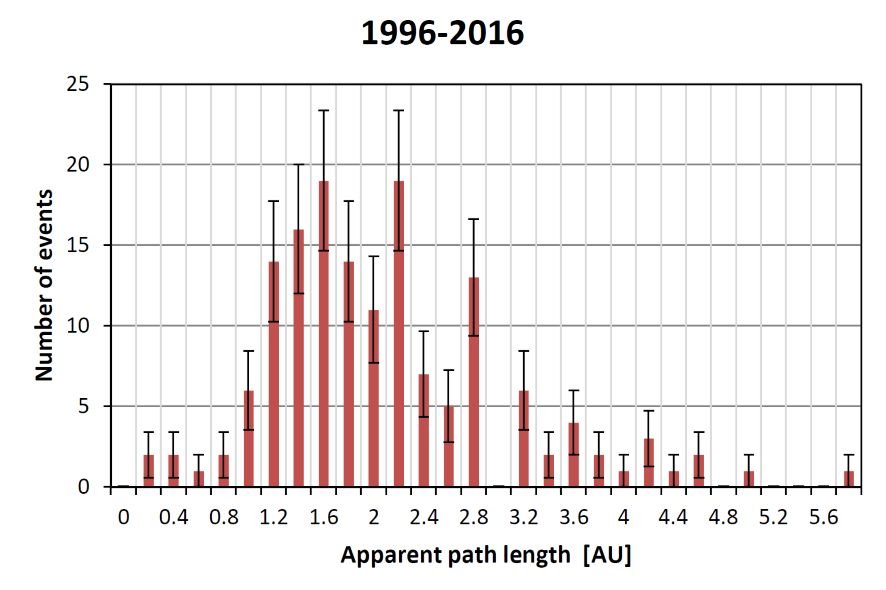}\\
\includegraphics[width=0.6\columnwidth]{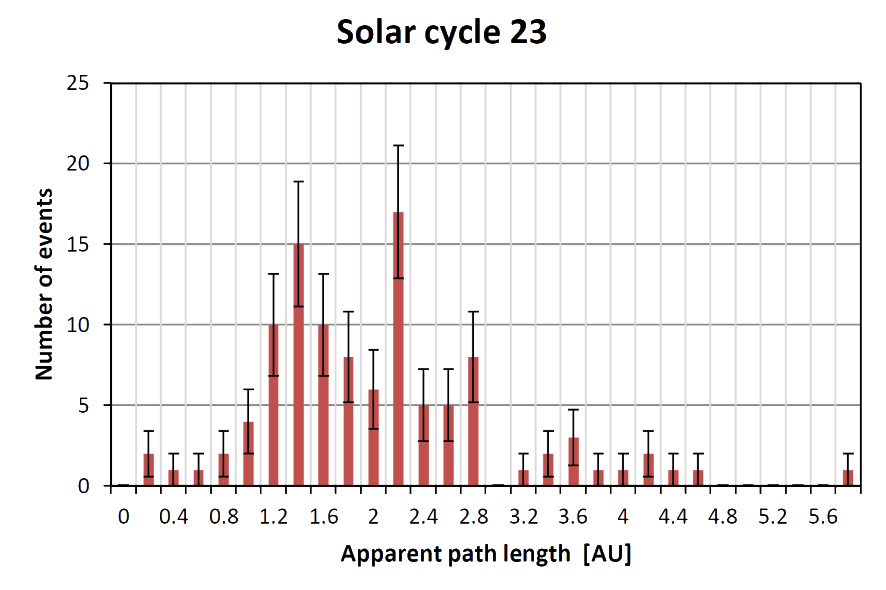}\\
\includegraphics[width=0.6\columnwidth]{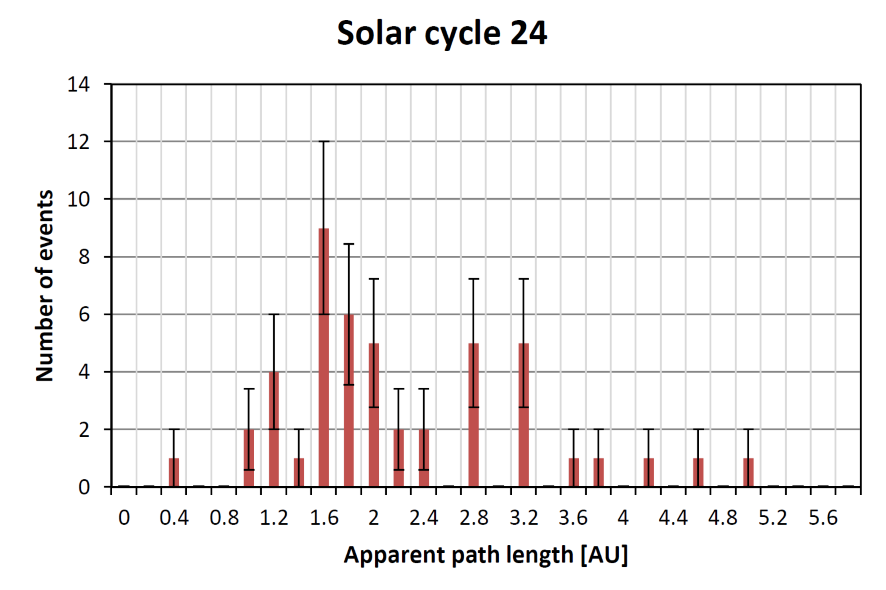}\\
\end{tabular}
\caption{\small Apparent path length distribution of the catalogued SEP events (161 in total). For seven events, the apparent path length was more than 6 AU; they are not shown here. The error bars denote the statistical error, and the values on the abscissa denote the lower limit of each path length bin.}
\label{fig1}
%\end{minipage}
\end{figure}

The distributions of apparent path lengths are shown in Figure \ref{fig1}. When all events are considered as one set, the distribution is somewhat reminiscent of a Gaussian one with its peak at $\sim$ 1.6 AU, but its  symmetry is broken by a second local maximum at $\sim$ 2.2 AU and a third, somewhat weaker one at $\sim$ 2.8 AU, as well as an apparent gap at 3.0 AU. The 1.6 AU peak is close to typical calculated values of the Parker spiral length. The corresponding 0.2 AU-wide bins contain 19, 19, and 13 events, respectively. Apart from the first one, these maxima may result from the poor statistics and they do not necessarily have any physical significance.

The 1996\textendash 2008 (solar cycle 23) and 2009\textendash 2016 (solar cycle 24) datasets, considered separately, present broadly similar distributions that would nevertheless seem to exhibit some difference in details. For the former, the first maximum occurs at 1.4 AU, and the third maximum at 2.8 AU is comparatively weak; the greatest number of events (17) falls into the 2.2 AU bin. In contrast, for the 2008\textendash 2016 events, the 1.6 AU bin dominates (9 events), and there is no clear maximum at 2.2 AU, while one possibly appears at 2.8 AU (and perhaps another at 3.2 AU). However, this set suffers from particularly poor statistics, and these features are unlikely to be of actual significance. For the same reason, the observations regarding the solar cycle 23 dataset must be considered with due caution. The average apparent path lengths (and their standard errors of the mean) are 2.47$\pm$0.20 AU for the 1996\textendash 2008 events, 2.60$\pm$0.22 AU for the 2009\textendash 2016 events, and 2.51$\pm$0.15 AU for the entire data set. The apparent path length usually exceeds the spiral field line length. 

These results are not in statistically significant disagreement with those reported by \citet{Vainio2013}. These authors reported a double-peaked distribution, with clear maxima at 1.4 AU and 2.2 AU, with possibly another very faint one at about 2.6 AU, while calling due attention to the poor statistics.

\begin{figure}[H]
\centering
\begin{tabular}{c}
\includegraphics[width=0.6\columnwidth]{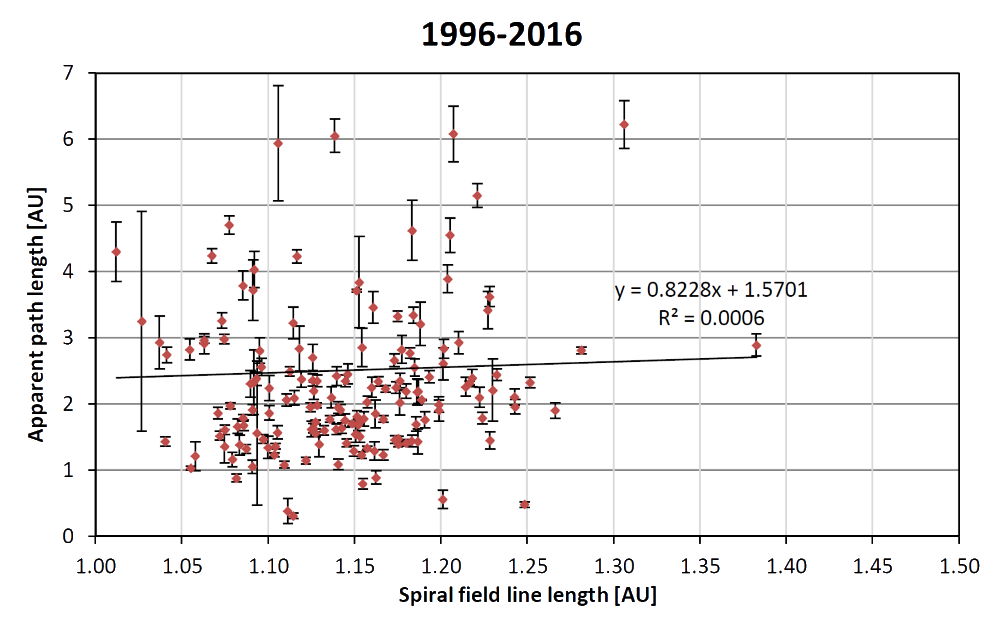}\\
\includegraphics[width=0.6\columnwidth]{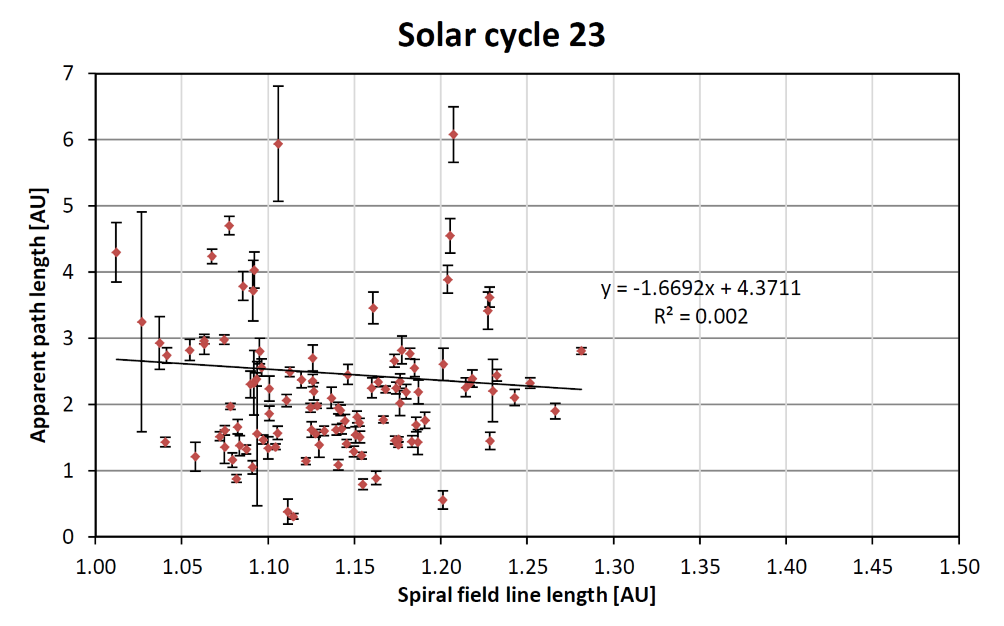}\\
\includegraphics[width=0.6\columnwidth]{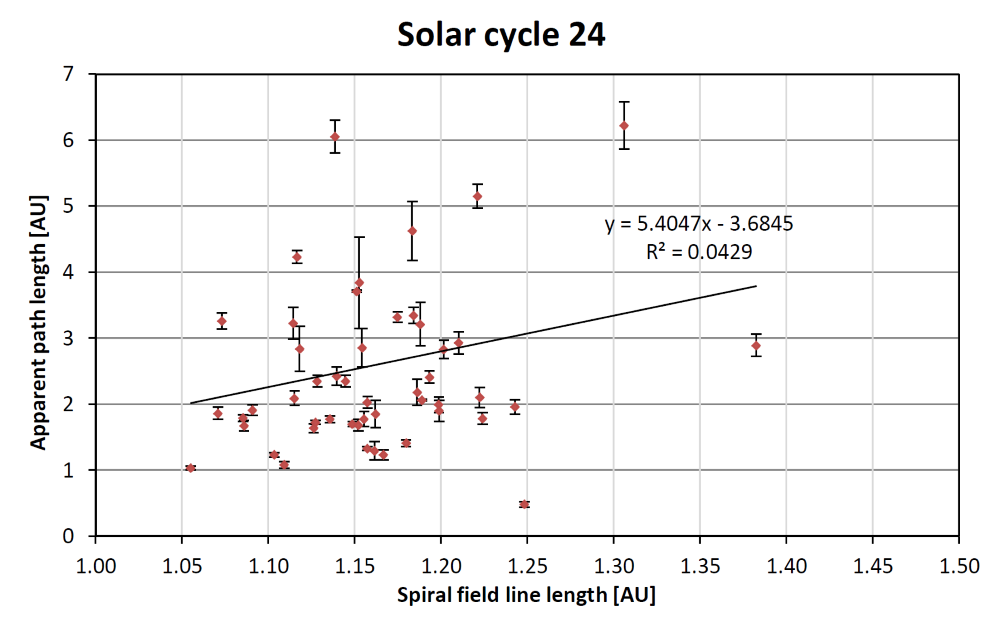}\\
\end{tabular}
\caption{\small The VDA apparent path length plotted against the spiral field line length for the catalogued events. The error bars were obtained from the VDA. Events for which apparent path length $>$ 7 AU are omitted from the subfigures but are included in the line fits.}
\label{fig2}
\end{figure}

The apparent path lengths are compared against the calculated spiral field line lengths in Figure \ref{fig2} and regression lines fitted to each data set using the least squares method to investigate whether or not a correlation can be said to exist between the two quantities. Even though the apparent path length is in the vast majority of cases greater than the spiral field line length\textemdash as is logical to expect for particles undergoing scattering\textemdash there is practically no correlation. When the subsets are considered separately, this observation becomes very clear, despite the fact that a fit to the whole data set would seem to produce a reasonable result, as the slope of the fit line has a value not very far removed from unity. A similar comparison was performed in \citet{Vainio2013}, where the fit line slope was also found to be near unity, possibly indicating a meaningful result. As the statistical significance of the fit was rather low, however, the authors suspected that this might be a fortuitous coincidence. When separate data sets for the solar cycles are considered, this indeed seems to be the case. This evidence suggests that the apparent path length is more dependent on other factors, such as the difficulties in onset determination caused by background intensity and the geometric factors of the ERNE particle telescopes not being constant with respect to incident particle energy, than on the spiral field line length.

As the final part of this subsection, we examine the distribution of the ratio of the spiral field line length and the apparent path length, $L$/$s$. This quantity can be used to arrive at an estimate for the average value of the pitch angle for the earliest-arriving particles. If we write

\begin{equation}
\mathrm{d} s = v \mathrm{d} t = \frac{v \mathrm{d} l}{v \mu} = \frac{\mathrm{d} l}{\mu},
\end{equation}
where d$l$ is the length element of the line connecting the particle source and observer and $\mu$ is the pitch angle cosine of the particles, and then integrate both sides of the equation, we see that

\begin{equation}
s = \frac{l}{\bar{\mu}},
\end{equation}
where $l$ is the actual distance along the field line between the source and the observer and $\bar{\mu}$ is the average value of $\mu$ for the earliest detected particles on their path from the Sun to the detecting spacecraft near the Earth. Taking $l = L$, we can express the effective value for $\mu$ as $\mu_{\rm eff} = \bar{\mu}$ = $L$/$s$.

\begin{figure}[H]
\centering
\includegraphics[width=0.6\columnwidth]{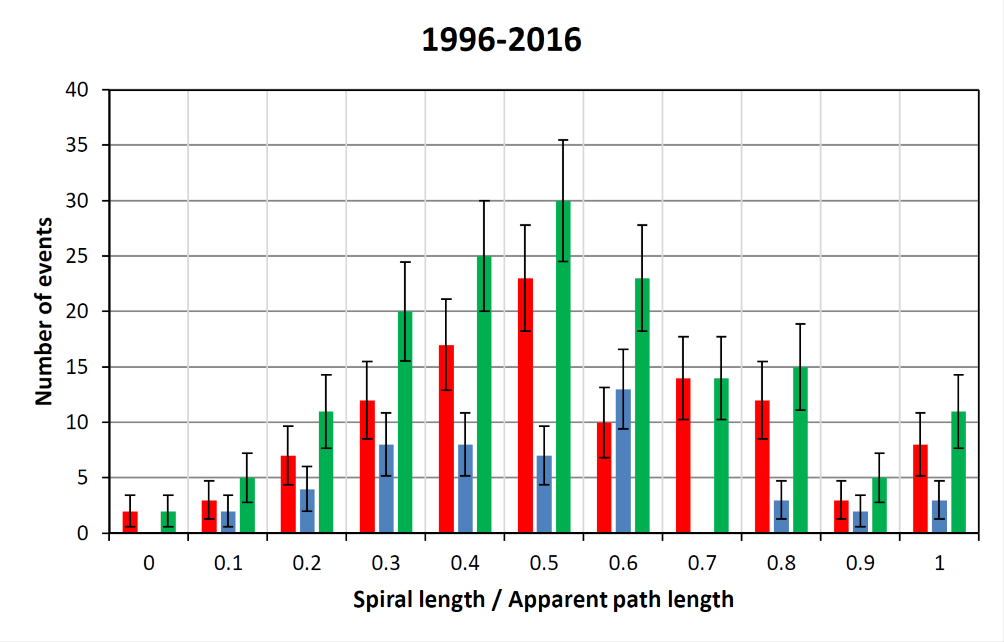}
\caption{\small The distribution of the ratio of the spiral field line length to the apparent path length obtained from VDA for the 1996\textendash 2016 events (161 in total). This quantity represents an estimate of the effective value of $\mu$ of the first arriving particles. The left-hand bar in each speed bin represents the solar cycle 23 (1996\textendash 2008) events, the middle bar the solar cycle 24 (2009\textendash 2016) events, and the right-hand bar all events during 1996\textendash 2016. The error bars denote the statistical error, only, and the numbers on the abscissa denote the lower limit of the $\mu$ bin. The last bin includes all the events for which $L$/$s$ $\geq$ 1.}
\label{fig13}
\end{figure}

Figure \ref{fig13} represents the distribution of $L$/$s$. Omitting the cases for which $L$/$s >$ 1 as unphysical, the mean and standard deviation, respectively, are for the datasets as follows: 0.54 and 0.20 (all events), 0.55 and 0.20 (the 1996\textendash 2008 events), and 0.52 and 0.20 (the 2009\textendash 2016 events). This result is in good agreement with the results given in \citet{Vainio2013}. It also underscores the fact, earlier found in simulation studies and pointed out by e.g. \citet{LintunenVainio2004} and \citet{Saiz2005}, that even the earliest-arriving particles cannot be regarded as moving directly along the magnetic field line, challenging one of the basic assumptions of VDA.

Overall, the mean and standard deviation values as such do not reveal any remarkable changes between the solar cycles. Despite the poor statistics of the 2009\textendash 2016 period, it would still appear that compared to solar cycle 23 as a whole, there have been very few events in 2009\textendash 2016 for which $L$/$s >$ 0.7; in other words, holding to the interpretation that the ratio $L$/$s$ indeed corresponds to the average pitch angle cosine, scattering tends to be more prominent in the cycle 24 events, during which time the Sun was much less active than in cycle 23. However, the difference between the distributions fails to meet the quantitative criteria for statistical significance\footnote{The statistical tests performed on the cycle 23 and cycle 24 data sets were Mann\textendash Whitney U test (\citealt{Mann-Whitney1947}) and two-sample Kolmogorov\textendash Smirnov test (in the latter, the cases in which $L$/$s >$ 1 were not considered). The  results of these tests are not shown here in detail.}, so no definite conclusions as to the differences between the two solar cycles can be made from this result alone.

\subsection{Flares and solar release times of particles}

The second main area of our statistical study was the investigation of particle release times derived with VDA for protons and TSA for electrons, alongside the X-ray signature of flares. With the time derivative maxima of the soft X-ray intensity known, it was possible to approximate the solar release times of relativistic electrons with the times of these maxima and compare them to TSA release times. We also made this comparison for proton VDA release times, and additionally considered the release time difference as a function of both flare longitude and the longitudinal distance of the flare from the Parker spiral footpoint.

Of the 128 listed flares, 45 (33 during 1996\textendash 2008, 12 during 2009\textendash 2016) belonged to the X class, 69 (45 during 
1996\textendash 2008, 24 during 2009\textendash 2016) to the M class, and 14 to the C class (seven during both 1996\textendash 2008 and 2009\textendash 2016). Flares that occurred over the western limb\textemdash i.e. those listed as "long. $\geq$ 90"\textemdash include one X class, ten M class, and two C class flares; they are included in the comparison of proton VDA release times and flare X-ray flux time derivative maxima (see below in this subsection) but are not considered in detail in the context of electron release times. The largest SEP-related flare in our catalogue was classified X20.0; it occurred on 2 April 2001 and has been a subject of study and discussion in its own right.\footnote{For instance, see \citet{Krucker2011} for a report on electromagnetic observations of this flare.} The flares associated with the 2009\textendash 2016 events are, on average, somewhat weaker than those associated with the 1996\textendash 2008 events: they constitute 27 \% of all X class flares, 35 \% of M class flares, and 50 \% of C class flares in Table 4, compared to the total share of some 34 \%.

\begin{figure}
\centering
\includegraphics[width=0.7\columnwidth]{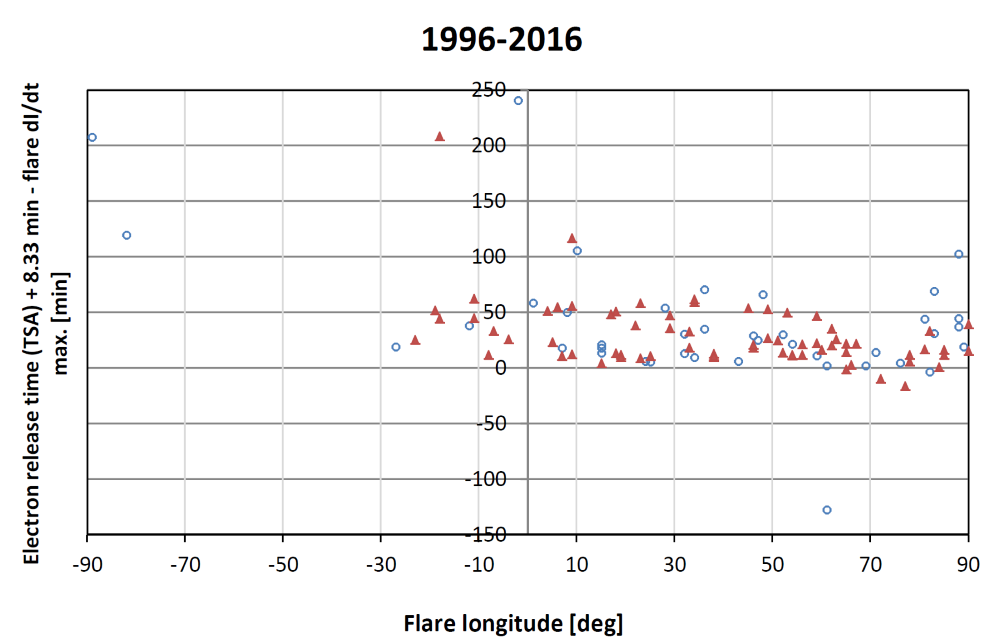}
\caption{\small Difference between apparent electron release time (TSA) and the time derivative maximum of X-ray intensity as a function of flare longitude for the 1996\textendash 2016 events. Solar cycle 23 (1996\textendash 2008) events are marked in red, solar cycle 24 (2009\textendash 2016) events in blue. Negative abscissa values refer to eastern and positive values to western longitude. Note the outlying data point near the lower right-hand corner.}
\label{fig16}
\end{figure}

Since the TSA calculation of electron release times, as employed in this work, does not take into account the magnetic connection between the particle acceleration site and the observer but instead assumes a direct path for the electrons along the Parker spiral, comparing the TSA release times with the maxima of the time derivative of soft X-ray intensity offers a simple means of studying approximately how well a given event is connected with the Earth. Figure \ref{fig16} demonstrates this by showing the difference between apparent electron release time and the maximum of the X-ray intensity time derivative as a function of flare longitude. Aside from a handful of outlying data points (these will be discussed further below), the flares are concentrated in the western longitudes, and for the great majority, the delay between the maximum of the X-ray intensity time derivative and the TSA-derived injection time is between zero and $\sim$ 60 minutes.

\begin{figure}
\centering
\includegraphics[width=0.7\columnwidth]{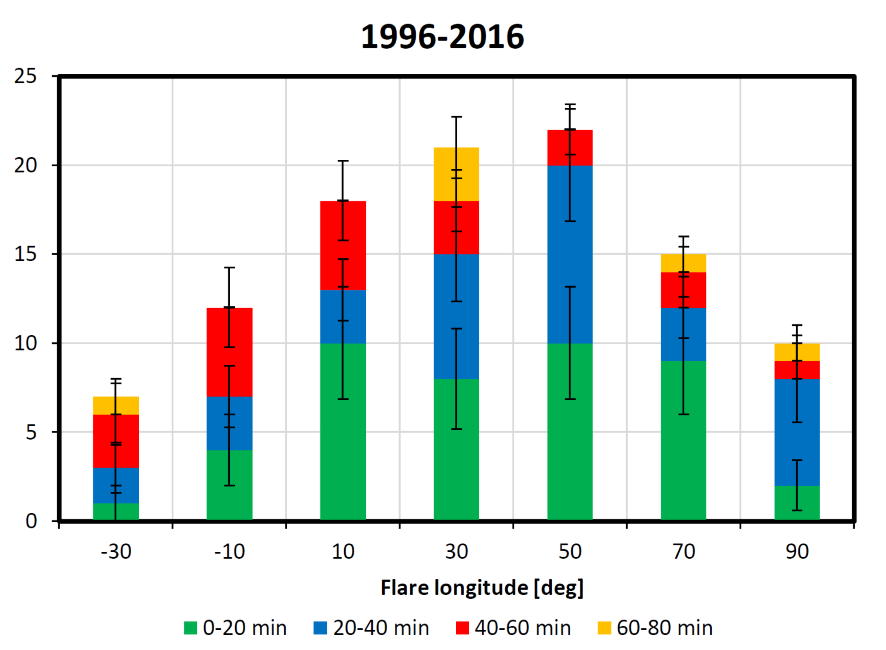}
\caption{\small The distribution of the 1996\textendash 2016 events with the location of origin more westerly than 30E and the difference between apparent electron release time (TSA) and the time derivative maximum of X-ray intensity within the range of [0,80] minutes. The error bars denote the statistical error, and the values on the abscissa (eastern longitudes expressed as negative, western longitudes as positive) denote the lower limit of each longitude bin. The last bin contains the western over-the-limb events.}
\label{fig17}
\end{figure}

Figure \ref{fig17} shows the 1996\textendash 2016 events for which the time delay is between 0 and 80 minutes and longitude is greater than 30E, with western over-the-limb events included, arranged into 20-degree longitude-wise and 20-minute time-wise bins. Considering only cases where the delay is between 0 and 20 minutes, the entire western solar hemisphere from 10W to near the limb is almost evenly populated: four successive 20-degree bins, starting from 10\textendash 30 degrees west, contain ten (six), eight (five), ten (seven), and nine (six) events for the entire data set (for solar cycle 23). If events with a delay between 0 and 40 minutes are selected, the longitude range of 50\textendash 70 degrees west is somewhat favoured over the others in both solar cycle 23 (15 cases) and the entire data set (20 cases). The flare locations in this longitude range are also expected to have the best magnetic connection with the Earth. The other two most populated bins are 10\textendash 30 degrees west with 13 (eight) events and 30\textendash 50 degrees west with 15 (eight) events for the entire data set (for solar cycle 23). In contrast, events for which the delay is between 40 and 60 minutes are proportionally more common in all the other bins than in the 50\textendash 70-degree one, which additionally does not contain any events with delay greater than 60 minutes. The statistics for the solar cycle 24 events, when considered separately, are too weak to allow any firm conclusions to be made, but the longitude distribution of the event-associated flares seems broadly similar in that set, as well.

However, when events originating from the eastern solar hemisphere are considered, a difference between the solar cycles does appear. For solar 
cycle 23, our listing includes only one case where the associated flare occurred at an eastern longitude greater than 20 degrees, namely event 58 on 
24 September 2001 (flare longitude 23E). In contrast, such events have been clearly more common and exhibited a generally greater angular separation 
from the nominal Earth-connected Parker spiral footpoint in solar cycle 24: event 127 on 22 September 2011 (89E), event 133 on 7 March 2012 (27E), 
event 151 on 25 October 2013 (73E), event 153 on 28 October 2013 (28E), and event 158 on 25 February 2014 (82E). If the cases where the flare onset 
precedes the proton event onset by more than six hours are also considered, event 41 on 21 January 2001 and event 81 on 6 September 2002 (probably 41E and 28E, respectively) were recorded for solar cycle 23, which are balanced by event 132 on 5 March 2012 and event 147 on 15 May 2013 (probably 54E and 
64E, respectively) during solar cycle 24. This observation is made more notable by the fact that the total number of 55\textendash 80 MeV proton events recorded by ERNE in cycle 24 is about 2/3 of that in cycle 23. (However, we note that lower-energy SEP events with a far-eastern location of origin are not very uncommon; see, e.g., \citealt{Cane2010}.)

Events with probable sources behind the eastern solar limb are quite rare in our listing, but they seem to follow a broadly similar pattern in that they have been at the very least equally frequent, if not more so, in solar cycle 24 than in solar cycle 23. While the timing of the particle injection causing event 66 (10 January 2002) is not certain (see discussion in \ref{ev_cat_2_6}), the suggested source longitude of 120E in \citet{Cane2010} may be possible; however, event 128 (3 November 2011) certainly originated at 152E, according to \citet{Richardson2014}. Event 163 (1 September 2014), with which no GOES flare could be associated with confidence and which will be considered by us in more detail in an upcoming study, may be another case originating over or behind the eastern limb.

A number of solar cycle 24 events in our catalogue apparently also originated from relatively far behind the western solar limb. Correlating our event listing with that of \citet{Richardson2014}, it appears that at least six events in 2009\textendash 2016 have probably been associated with sources at longitudes of $\geq$ 130 degrees, one (that of event 146 on 24 April 2013) being as distant as 175W. Unfortunately, similar coverage of solar farside flare activity is not available for events prior to the beginning of the STEREO mission in late 2006, so a thorough comparison between the two cycles in this respect is at best difficult.

So, while the overall relatively poor statistics compel a cautious approach to making definite conclusions, the evidence nonetheless points to solar cycle 24 being a more favourable period than solar cycle 23 for high-energy events to be detected when the particles involved originate from solar locations that are usually not well magnetically connected with the near-Earth observer. This, in turn, might possibly imply that very wide SEP events have been more characteristic of cycle 24 than cycle 23, but our data set alone is not sufficient to establish such a result.

\begin{figure}[H]
\centering
\includegraphics[width=1.0\columnwidth]{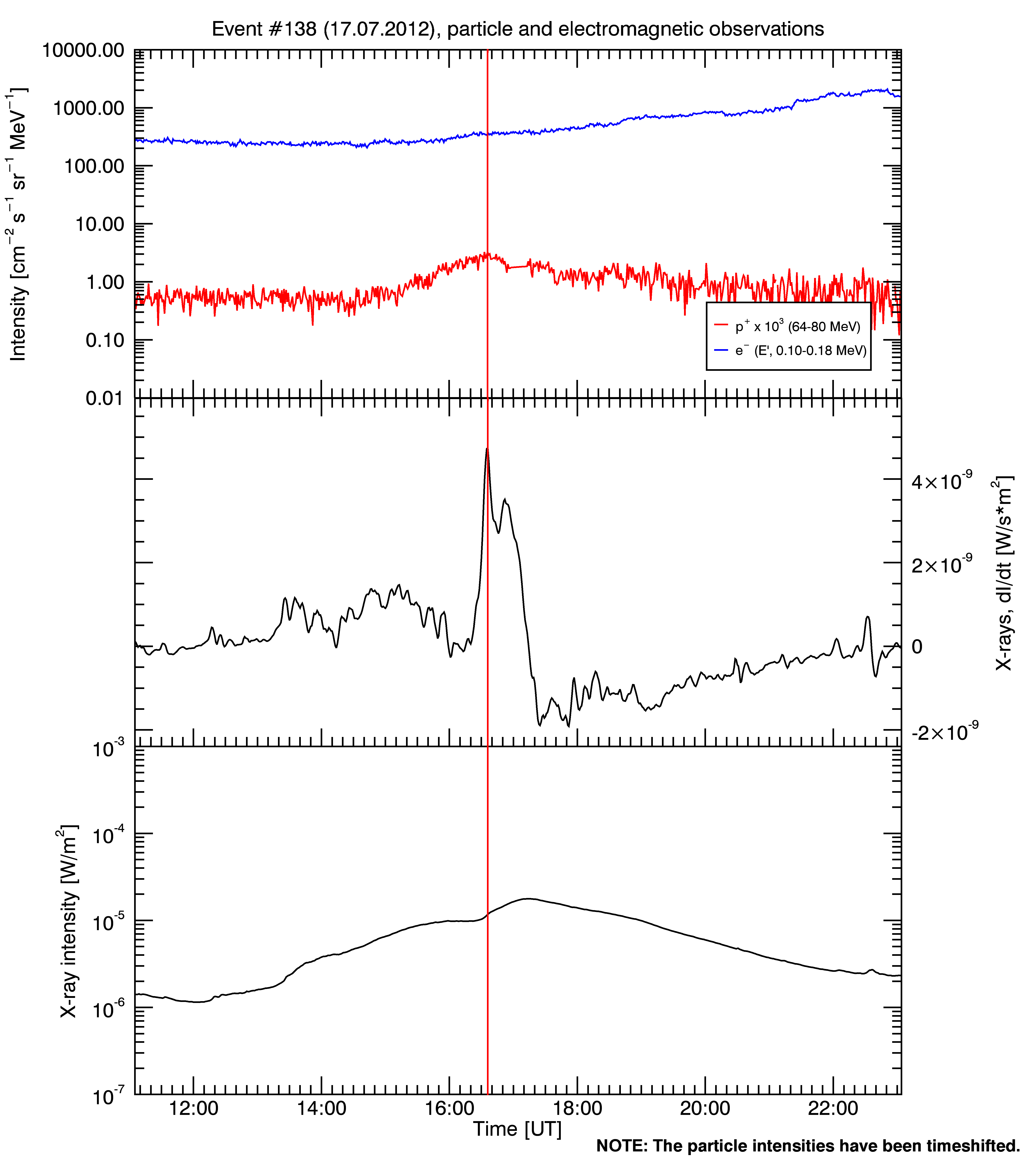}
\caption{\small The particle and X-ray observations related to the SEP event of 17 July 2012 (event 138). Shown here are proton and electron intensities (top panel, the former multiplied by 10$^3$), the time derivative of the soft X-ray intensity (middle panel) and the soft X-ray intensity (bottom panel). The moment of the maximum of d$I$/d$t$ for X-ray intensity is marked with a vertical red line. The particle intensities have been backshifted in time by ($L$/$v$ - 500) seconds, where $L$ = 1.13 AU and $v$ = the mean speed of the particle species of interest.}
\label{fig_supp1}
\end{figure}

In two events, the X-ray intensity exhibits such a slow rise that its time derivative maximum occurs after the arrival of the first protons at the Earth. The most striking example is the SEP event of 17 July 2012: an M1.7 class flare begins at 12:03, ACE/EPAM detects the onset of an electron event at 14:33, followed by the detection of the onset of a 67.7-MeV proton event by SOHO/ERNE at 15:05. However, the time derivative of the X-ray intensity does not reach its maximum until 16:36, more than four and a half hours after the onset of the flare. This event can be seen as an outlier near the bottom right-hand corner in Figure \ref{fig16}. It is obvious that the electron release must have begun some time between the flare and the electron event onsets, probably close to the TSA-derived (actual) release time of 14:20. Figure \ref{fig_supp1} shows the event in question as observed in proton (SOHO/ERNE), electron (ACE/EPAM), and X-ray (GOES) intensity, with a common time axis. The extremely slow rise of the latter is quite pronounced here. However, it should be noted that there is very low-intensity X-ray activity showing peaks in the time derivative of X-ray intensity also earlier than the identified peak, related to the global X-ray maximum of the flare. This could signal the acceleration and release of the first-observed particles in the event. Somewhat similar but much less conspicuous cases with probable particle injections and accompanying considerable X-ray activity preceding the main event-associated flare, consequently resulting in a "too early" particle onset, are event 69 (20 February 2002) for both protons and electrons, as well as events 38 (8 November 2000) and 76 (14 August 2002) for electrons.

On the other hand, the outlying points for which the difference in time between the estimated electron release and the flare X-ray intensity reaching its maximum slope is positive and large, aside from the far-eastern events discussed earlier, very likely result from insufficient sensitivity of the electron event onset determination or a high pre-event electron background. These tend to lead to delayed detection of the onset.

Approximating the solar rotation as rigid, we can estimate the longitudinal distance of the flare and the footpoint of the Parker spiral that connects the solar surface to the Earth at the SEP event onset, often called the connection angle, as follows:

\begin{equation}
\Delta \phi = \lvert \phi _{\rm flare} - \omega_{\odot} \frac{r_{\rm SC}}{u_{\rm SW}}\rvert.
\end{equation}

\begin{figure}
\centering
\includegraphics[width=0.7\columnwidth]{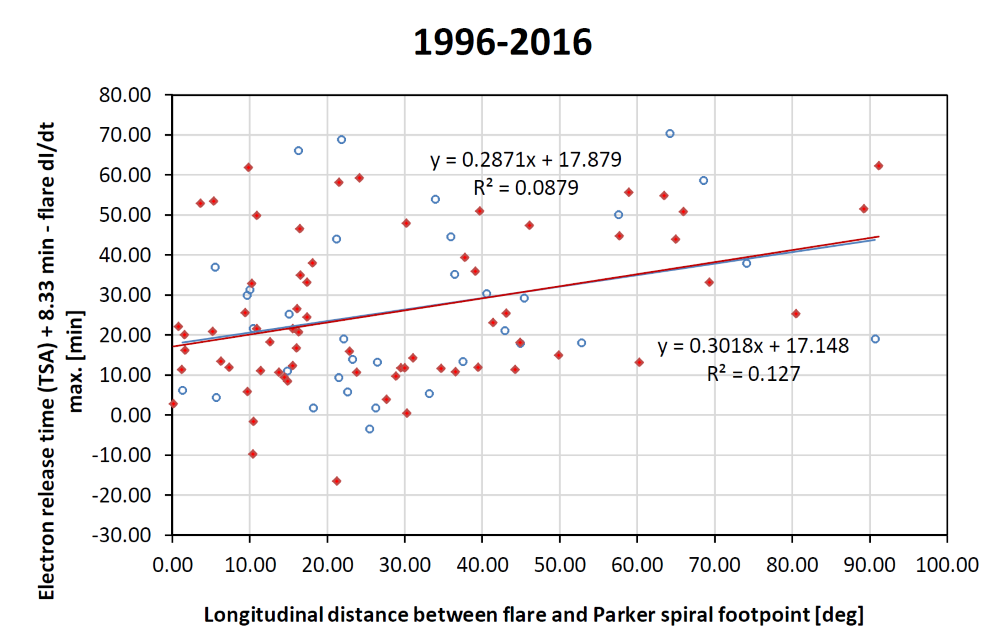}
\caption{\small Difference between apparent electron release time (TSA) and the time derivative maximum of X-ray intensity as a function of the connection angle (the longitudinal distance of the flare from the Earth-connected Parker spiral footpoint) for the catalogued events, with outlying data points (see text) omitted. The red symbols represent solar cycle 23 (1996\textendash 2008) events, and the blue symbols represent solar cycle 24 (2009\textendash 2016) events; the fit lines coincide nearly perfectly in this graphic. The lower equations refer to the linear fit for solar cycle 23 events, the upper ones to the linear fit for solar cycle 24 events.}
\label{fig19}
\end{figure}

The time difference between the TSA-estimated electron release and the time derivative maximum of the X-ray intensity is plotted as a function of $\Delta \phi$ in Figure \ref{fig19}. Note that the outlying data points, for which the connection angle is more than 100 degrees or the absolute value of the time difference is more than 100 minutes, are omitted from this analysis. Again, the trend is that the better connected a flare/SEP release region is, the shorter is the delay between the electron release time as estimated with TSA and the X-ray intensity time derivative maximum. To investigate quantitatively the level of correlation between the longitudinal distance and the time difference, a regression line was fitted to both solar cycle 23 and 24 data sets. When only the core population, shown in Figure \ref{fig19}, is considered in the fit, the results are similar for both cycles. While the statistical significance of the fit is not particularly great in either of these instances, and a lot of scatter is present in the data, the slopes and the intercepts of the lines are nevertheless sensible in that both are greater than zero and the latter lie in the range of 15\textendash 20 minutes. This indicates that in a very well connected event, the maximum energy release in the flare/CME, as estimated from the X-ray intensity time derivative maximum\footnote{The relationship between CME kinematics and X-ray intensity profiles is discussed in e.g. \citet{Zhang2006}.}, would tend to occur some twenty minutes or less before the electron release, as estimated from TSA. Such a result appears reasonable, and it also suggests that using the time derivative maximum of the flare X-ray intensity is a basically valid method for approximating the time of global particle release.

\begin{figure}[H]
\centering
\includegraphics[width=0.7\columnwidth]{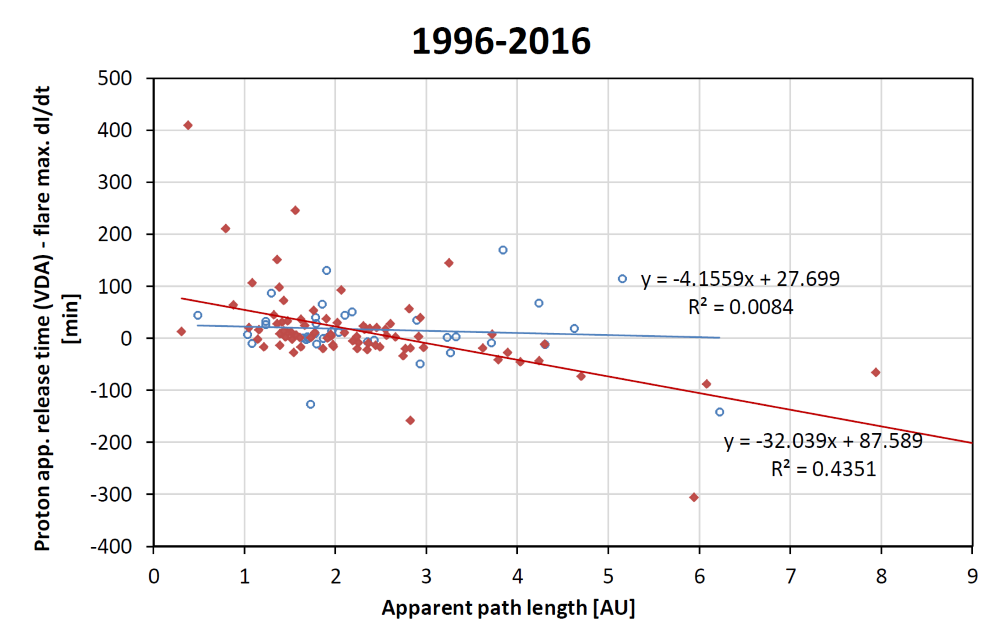}
\caption{\small Difference of apparent proton release time (as obtained from VDA) and the time derivative maximum of the soft X-ray intensity as a function of VDA apparent path length $s$ for the catalogued events. The red symbols represent solar cycle 23 (1996\textendash 2008) events, and the blue symbols represent solar cycle 24 (2009\textendash 2016) events. The lower equations refer to the linear fit for solar cycle 23 events, the upper ones to the linear fit for solar cycle 24 events. One event, for which $s >$ 13 AU, is not shown here.}
\label{fig22}
\end{figure}

\begin{figure}[H]
\centering
\includegraphics[width=0.7\columnwidth]{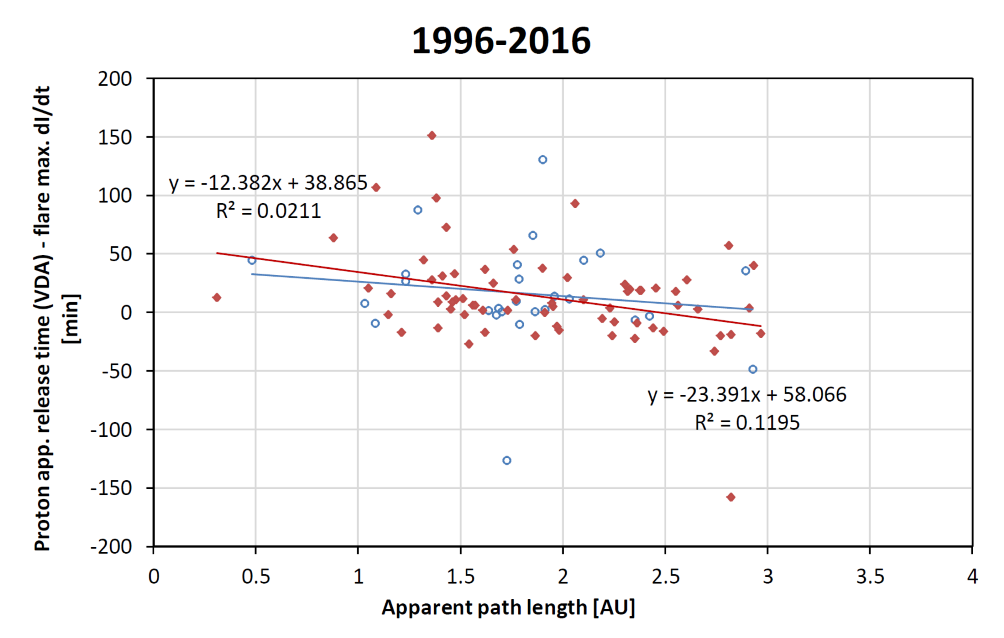}
\caption{\small Difference of apparent proton release time (as obtained from VDA) and the time derivative maximum of the soft X-ray intensity as a function of VDA apparent path length $s$ for the catalogued events, with all data points that have $s >$ 3 AU or the absolute value of the time difference in excess of 200 minutes omitted. The red symbols represent solar cycle 23 (1996\textendash 2008) events, and the blue symbols represent solar cycle 24 (2009\textendash 2016) events. The lower equations refer to the linear fit for solar cycle 23 events, the upper ones to the linear fit for solar cycle 24 events.}
\label{fig23}
\end{figure}

Figures \ref{fig22} and \ref{fig23} show the difference between apparent VDA proton release time and the time derivative maximum of soft X-ray intensity as a function of apparent path length; in the latter, the events for which the path length is greater than 3.0 AU or the absolute value of the time difference is greater than 200 minutes are omitted from the analysis. Linear fits to the plotted data points are also shown in these Figures. Again, the points are strongly scattered, and the linear fits are not very significant statistically. It is seen that the events which most likely have a reasonable path length ($s$ falls between 1 AU and some 3 AU) also have fairly small release time differences only weakly dependent on the path length. In all considered data sets, this time difference tends to become increasingly negative as $s$ increases. A plausible explanation is that VDA produces over-estimated path lengths particularly for events with large $s$ and consequently proton release times that are too early.

\subsection{SEP event-associated CMEs}

As the final part of our statistical analysis, we briefly examined the CMEs associated with the 55\textendash 80 MeV proton/flare events. The width and speed of the CMEs, as indicators of their kinetic energy and size, were compared between the solar cycles.

The relationship between fast CMEs and SEP events has long been recognized. For instance, Reames (\citealt{Reames1999} and references therein) pointed out the essential nature of CME speed for particle acceleration in gradual SEP events, and \citet{Kahler2001} derived a correlation between CME speed and SEP event peak intensity at the Earth. More recently, \citet{Dierckxsens2015} have estimated the mean speed of CMEs associated with  SEP events (ranging in proton energy from a few MeV upwards) to be about 1000 km/s, based on the SOHO LASCO CME Catalog data. While noting that for halo CMEs, the likelihood of an SEP event occurring is more dependent on the magnitude of the CME-associated flare than CME speed, they report that SEP event occurrence probability increases for all CME widths as CME speed increases and that halo CMEs are significantly more likely to be associated with SEP events than non-halo CMEs, as defined in the SOHO LASCO CME Catalog.

We first considered the angular sizes and speeds of the SEP-related CMEs listed in our catalogue. Omitting the seven cases with an uncertain size estimate, halo CMEs (extending a full 360 degrees around the disk of the Sun as seen from the Earth) constitute 72 of 101 cases (72 \%) for the 1996\textendash 2008 events, 48 of 59 cases (82 \%) for the 2009\textendash 2016 events, and 120 of 160 for all events with a well documented CME (75 \%); this result again underscores the fact that halo CMEs are typical for SEP events. The average speed for the 1996\textendash 2008 CMEs was 1361 km/s, 1276 km/s for the 2009\textendash 2016 events, and 1331 km/s for all events. These results are in good agreement with those reported by \citet{Papaioannou2016}, where an average speed of some 1390 km/s is given for SEP-related CMEs. The authors point out that most CMEs associated with SEP events are faster than 1000 km/s but that there are some slow ($\leq$ 500 km/s) SEP-related CMEs, as well; again, our work confirms these observations. As the formation of a strong shock, facilitated by high CME speed, is thought to be essential for particle acceleration, slow CMEs associated with SEP events might warrant further study.

\begin{figure}
\centering
\includegraphics[width=0.6\columnwidth]{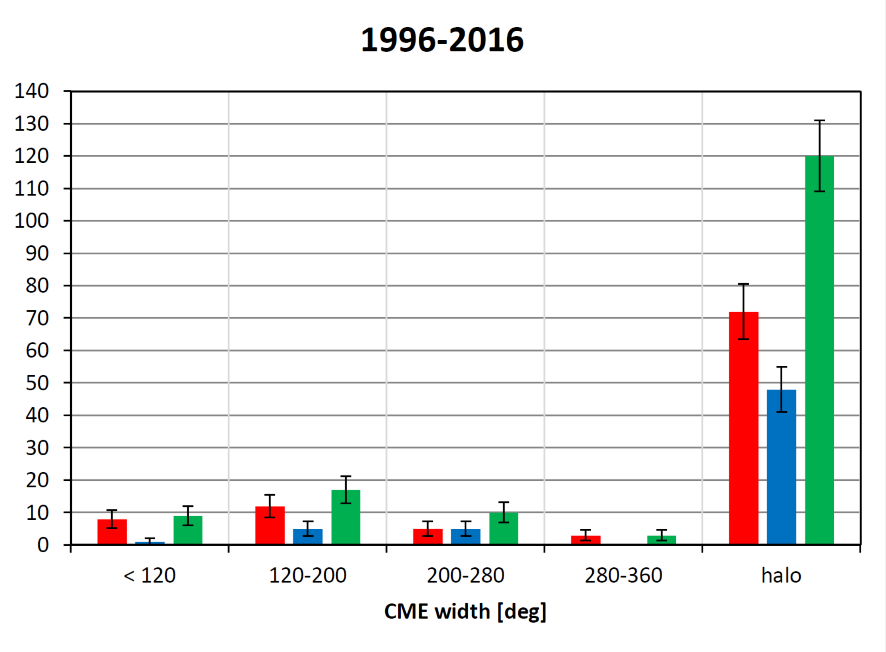}
\caption{\small The angular sizes of SEP-associated CMEs for the catalogued events. The left-hand bar in each speed bin represents the solar cycle 23 (1996\textendash 2008) events, the middle bar the solar cycle 24 (2009\textendash 2016) events, and the right-hand bar all events during 1996\textendash 2016. The error bars denote the statistical error. Seven CMEs, for which only an upper limit estimate of the angular size is listed, are omitted.}
\label{fig25}
\end{figure}

\begin{figure}
\centering
\includegraphics[width=0.6\columnwidth]{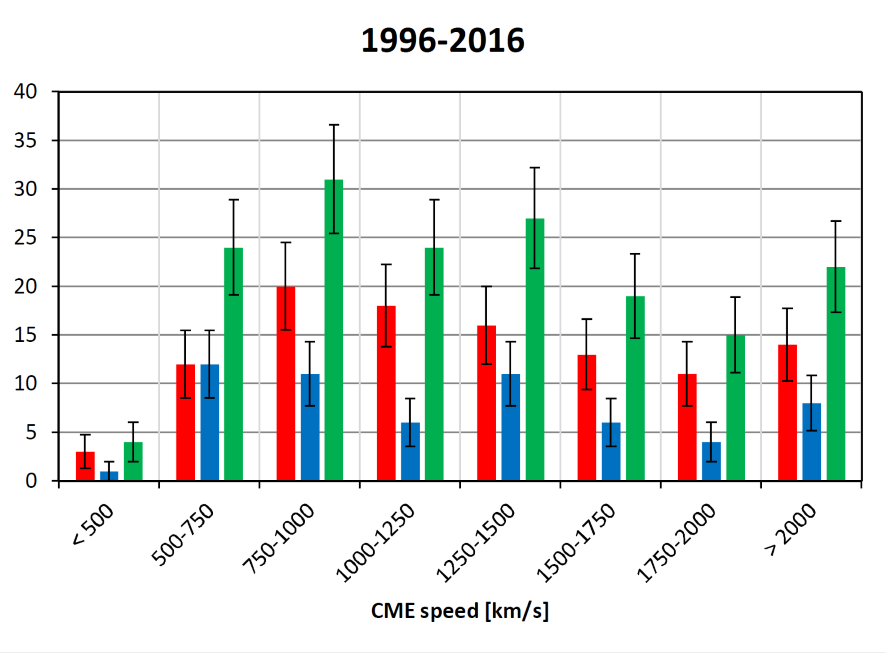}
\caption{\small The distribution of speeds of SEP-associated CMEs for the catalogued events. The left-hand bar in each speed bin represents the solar cycle 23 (1996\textendash 2008) events, the middle bar the solar cycle 24 (2009\textendash 2016) events, and the right-hand bar all events during 1996\textendash 2016. The error bars denote the statistical error.}
\label{fig26}
\end{figure}

Figure \ref{fig25} represents the distribution of the angular size of the CMEs, and Figure \ref{fig26} the distribution of their speed. The dominance of halo CMEs in both periods of interest is obvious. There appears to be a mild preference for what might be termed "medium-speed" SEP event-associated CMEs (750\textendash 1000 km/s) and an expected clear under-representation of slow ($<$ 500 km/s) CMEs as compared to CMEs in general, but overall the speed distributions of SEP-associated CMEs seem to possess no surprising features. Again, the comparative analysis of the two solar cycles is hampered by the poor statistics of the 2009\textendash 2016 period, for which the most populated bin contains a mere 12 events, but it would appear that no truly significant differences exist between the earlier and the later data sets. This conclusion essentially seems to hold for the size distribution, as well: aside from an increased proportion of halo events, no remarkable differences were found between the two time periods.

\citet{Gopalswamy_etal_2015} report that while halo CMEs were equally abundant in both solar cycles 23 and 24, the sources of those of the latter cycle were more uniformly distributed in longitude than those of the previous cycle, which is ascribed to the heliospheric pressure (the sum of plasma and magnetic pressures) being on average lower during cycle 24 than during cycle 23, allowing CMEs to grow into full halo events more often. Thus, it would seem that this variation also has implications to SEP events, bearing in mind the importance of the CME-induced shocks for particle acceleration (\citealt{Reames1999}): particles originating from sites magnetically poorly connected to the observer are more effectively accelerated and observed as an SEP event near the Earth when the associated CME is able to expand against a lower-than-usual ambient pressure, attaining a great width and thereby forming a shock also on the Earth-connected field line more easily.

However, it is noteworthy that the available evidence points to a clear decrease of SEP events with hard spectrum during solar cycle 24, as compared to cycle 23 (see, e.g., \citealt{Mewaldt2015} for an in-depth discussion). In our data, the seven-year period from 1996 to 2003 contains 92 proton events in the 55\textendash 80 MeV energy range, while only 62 are recorded for 2009\textendash 2016. (Both of these figures probably somewhat underestimate the respective numbers of actual SEP events, owing to gaps in the ERNE data.) Furthermore, only one ground level enhancement (GLE) event has so far occurred during solar cycle 24, whereas there were 13 in the first eight years (1996\textendash 2003) of solar cycle 23 (as per the GLE database, \url{http://gle.oulu.fi/}); the corresponding numbers of sub-GLE events\footnote{Sub-GLEs are energetic particle events extending beyond $\sim$ 300 MeV in energy but showing no reliably detectable signal on ground level.} for these time periods are 11 and 10 (Vainio et al. 2017; submitted manuscript). On the other hand, \citet{Chandra2013}, who considered SEP events associated with type II radio bursts detected during the rising phases of both solar cycles (1996\textendash 1998 and 2009\textendash 2011, respectively), found that there were similar numbers of events on the $>$ 10 MeV GOES proton channel (14 events with maximum intensity greater than 1 pfu for both time periods).

\section{Conclusions and outlook}
\label{conclusions}
Studying SOHO/ERNE data, we identified 176 55\textendash 80 MeV proton events that occurred during 1996\textendash 2016. We determined their onset times at 1 AU, their maximum intensities, proton and oxygen fluences, and Fe/O ratio, and performed VDA and TSA for the events. The associated energetic electron events were also identified from ACE/EPAM data and their onsets determined. TSA was then performed for the electron events. Lastly, we attempted to identify X-ray flares and CMEs corresponding to the particle events, obtaining estimates for particle solar release times from the maxima of the time derivative of the soft X-ray intensity. This accumulated information is presented in the form of an event catalogue, in two parts (Tables 3 and 4).

In all but a few cases, the identification and subsequent analysis of the electron event accompanying the proton event was successful. An associated CME could be identified for practically all of the SEP events considered by us, and the majority were also associated with a flare of NOAA classification C1.0 or greater, occurring on the visible solar disk and less than six hours before the onset of the proton event near the Earth.

We performed a statistical analysis on the VDA and TSA path length results, both for the entire event set as a whole and in terms of comparison between the solar cycles 23 and 24. In addition, the SEP-associated X-ray flares and CMEs were briefly investigated. Comparison of the two solar cycles was made 
somewhat difficult by the poor statistics of especially the latter cycle, but some conclusions can nevertheless be made. For the most part, our results are consistent with those presented in \citet{Vainio2013}, a previous study involving many of the same topics of interest.

We found that proton VDA produces reasonable results\textemdash for the apparent path length $s$ and the release time $t_0$\textemdash for most 
events, demonstrating that its underlying assumptions also appear to be reasonable, even if not absolutely correct. However, the results must be regarded as dubious in cases where the apparent path length is less than 1 AU or considerably more than about 3 AU. Provided that the apparent path length is within these realistic limits, there is no significant correlation between the apparent path length and release time $t_0$, or between the spiral field line length $L$ and the apparent path length $s$, as derived from VDA for protons. It appears likely, therefore, that $s$ may be more affected by such conditions as the background intensities and the energy dependence of the geometric factors of the ERNE particle telescopes than the spiral field line length. As established in previous studies, these protons cannot be regarded as propagating without substantial scattering in the interplanetary medium. In fact, the average apparent path length for the first-arriving particles exceeds the spiral field line length by a factor of about two. Furthermore, scattering of protons tended to be slightly more pronounced during solar cycle 24 (average $L$/$s \approx$ 0.52) than during cycle 23 (average $L$/$s \approx$ 0.55), but this apparent increase  does not meet the criteria for statistical significance and thus cannot be considered a meaningful indication of difference between the solar cycles at present. A greater number of data points\textemdash obtainable by increasing the energy range to cover, for instance, all $>$ 25 MeV proton events\textemdash is needed to decide whether the two cycles might indeed differ in this respect.

Other essential conclusions of our study can be summarized as follows:

\begin{enumerate}[i.)]
\item Allowing for inaccuracies in determining the electron event onset, the 
maximum of time derivative of soft X-ray intensity appears to be a valid indicator for the global solar release time of energetic electrons, as evidenced by the time difference between it and TSA-derived moment of electron solar release. As expected, the least values of the time difference tend to occur for the events that, judging by the solar longitude of the associated flare, have the best magnetic connection with the observer. However, a considerable amount of scatter is present in the data.
\item SEP events with a considerable longitudinal distance between the associated flare and the footpoint of the Sun\textendash Earth magnetic line 
were noticeably more frequent in cycle 24 than in cycle 23; for three well-documented events in our catalogue, the solar longitude of the flare was more than some 60 degrees east, while no directly comparable case occurred in cycle 23. High-energy events with a far-eastern origin in cycle 24 also generally seem to outnumber similar events in cycle 23. This raises the possibility of an increase in angular width for these events, which could be a feature of the current cycle.
\item The majority of CMEs associated with SEP events were of the halo type in both cycles, but their proportion was greater during cycle 24. This 
can be attributed to the lower total pressure in the heliosphere, allowing more CMEs to become halo events. This observation is in concurrence with a  result previously reported by \citet{Gopalswamy_etal_2015}. As the lower total pressure also facilitates the forming of a shock\footnote{This interpretation assumes similar mass densities in the heliosphere during both solar cycles in question.}, the CMEs of cycle 24 may have been more efficient in accelerating SEPs towards the Earth from poorly connected sites than those of cycle 23, accounting together with ii.) for the SEP events originating from locations with a poor expected magnetic connection to the observer. However, events with very high particle energies have been generally rare in cycle 24.
\end{enumerate}

It would be highly desirable to extend the study to cover the rest of solar cycle 24 as soon as it comes to an end. A comprehensive data coverage of several more years could improve the statistics and would also allow direct comparison between the respective rising and falling phases of the two cycles. Another potentially fruitful line of investigation might be to include energetic particle data from other spacecraft, such as the STEREO probes, so as to obtain a more comprehensive spatial coverage of the SEP events. However, a study of this sort would be practically limited to cover only cycle 24.

The results detailed in our work may be freely utilized in future studies. An online version of the catalogue presented in this paper is available at \url{http://swe.ssa.esa.int/web/guest/utu-srl-federated}; in addition to the tabulated information given here, it offers plots of various observable quantities over the events.

\begin{acknowledgements}
The research described in this paper was supported by ESA contract 4000113187/15/D/MRP. We would like to acknowledge and express our gratitude to the organizations and teams responsible for maintaining the data sources used in this article (SEPServer, SOHO LASCO CME Catalog, SolarSoft Latest Events Archive, NOAA/Solar-Terrestrial Physics at the National Centers for Environmental Information, Coordinated Data Analysis Web, the ACE Science Center, and the GLE database). SOHO LASCO CME Catalog: this CME catalog is generated and maintained at the CDAW Data Center by NASA and The Catholic University of America in cooperation with the Naval Research Laboratory. SOHO is a project of international cooperation between ESA and NASA.

O.R. wishes to thank the Vilho, Yrj\"o and Kalle V\"ais\"al\"a foundation for financial support.

A.P. would like to acknowledge gratefully the invitation by R.V. and the support of the University of Turku (UTU) that made his working visit to UTU possible.

The editor thanks Ian Richardson and an anonymous referee for their assistance in evaluating this paper.
\end{acknowledgements}

%% The bibliography (.bib) file:

\bibliography{proton_cat_bib}

%\end{linenumbers}

\end{document}